
\input epsf \def\figin#1{#1}

\input harvmac

\def\s{\sigma}
\def\e{\eta}
\def\k{\kappa}
\def\ee{{\rm e}^}
\def\d{{\rm d}}
\def\IR{\relax{\rm I\kern-.18em R}}
\font\cmss=cmss10 \font\cmsss=cmss10 at 7pt
\def\IZ{\relax\ifmmode\mathchoice
{\hbox{\cmss Z\kern-.4em Z}}{\hbox{\cmss Z\kern-.4em Z}}
{\lower.9pt\hbox{\cmsss Z\kern-.4em Z}}
{\lower1.2pt\hbox{\cmsss Z\kern-.4em Z}}\else{\cmss Z\kern-.4em Z}\fi}
\def\M{\Lambda}
\def\one{{\bf 1}}
\def\delbar{\overline\del}
\def\inbar{\vrule height1.5ex width.4pt depth0pt}
\def\IC{\relax\thinspace\hbox{$\inbar\kern-.3em{\rm C}$}}
\def\figinsert#1#2#3{\topinsert\figin{#2}\centerline{\vbox{\baselineskip12pt
\advance\hsize by -1truein\noindent\footnotefont{\bf Fig.~\xfig#1:} #3}}
\bigskip\endinsert}

\Title{\vbox{\baselineskip12pt\hbox{hepth@xxx/9202092}\hbox{LA-UR-92-640}%
\hbox{NEIP92-001}}}
{\vbox{\centerline{Strings on Curved Spacetimes:}
   \vskip2pt\centerline{Black Holes, Torsion, and Duality}}}

\centerline{Paul Ginsparg\footnote{$^1$}{email: ginsparg@xxx.lanl.gov} and
Fernando Quevedo\footnote{$^2$}
{Present address: Institut de Physique, Universit\'e de Neuch\^atel,
 Rue A.-L. Br\'eguet 1 CH-2000, Switzerland. email: phquevedo@cnedcu51.bitnet.
Supported by Swiss National Foundation.}}

\bigskip\centerline{Los Alamos National Laboratory}
\centerline{Theoretical Division, MS-B285}\centerline{Los Alamos NM \ 87545}

\vskip .3in
\noindent We present a general discussion of strings propagating on
noncompact coset spaces $G/H$ in terms of gauged WZW models, emphasizing
the role played by isometries in the existence of target space duality.
Fixed points of the gauged transformations induce metric singularities
and, in the case of abelian subgroups $H$, become horizons in a dual
geometry. We also give a classification of models with a single timelike
coordinate together with an explicit list for dimensions $D\leq 10$. We
study in detail the class of models described by the cosets
$SL(2,\IR)\otimes SO(1,1)^{D-2}/SO(1,1)$. For $D\geq 2$ each coset
represents two different spacetime geometries: (2D black hole)$\otimes
\IR^{D-2}$ and (3D black string)$\otimes \IR^{D-3}$ with nonvanishing
torsion. They are shown to be dual in such a way that the singularity of
the former geometry (which is not due to a fixed point) is mapped to a
regular surface (i.e.\ not even a horizon) in the latter . These cosets
also lead to the conformal field theory description of known and new
cosmological string models.

\Date{2/92}

\noblackbox

\lref\bn{I. Bars and D. Nemechansky, Nucl. Phys. B348 (1991) 89.}
\lref\witten{E. Witten, Phys. Rev. D44 (1991) 314.}
\lref\rwzw{E. Witten, Comm. Math. Phys. 92 (1984) 455.\semi
For a review see P. Goddard and D. Olive, Int. J. Mod. Phys. 1 (1986) 303.}
\lref\wzwg{E. Witten, Nucl. Phys. B223 (1983) 422\semi
K. Bardacki, E. Rabinovici and B. Saering, Nucl. Phys. B301 (1988) 151 \semi
D. Karabali and H. J. Schnitzer, Nucl. Phys. B329 (1990) 649, and references
therein.}
\lref\busc{T. Buscher, Phys. Lett. 194B (1987) 59;
 Phys. Lett. 201B (1988) 466.}
\lref\vhol{For a review see, J. W. van Holten, Z. Phys. C27:57 (1985).}
\lref\gmv{M Gasperini, J. Maharana and G. Veneziano, CERN preprint
CERN-TH-6214/91 (1991).}
\lref\sen{A. Sen, preprints  TIFR/TH/91-35 and TIFR/TH/91-37\semi
S. Hassan and A. Sen, preprint TIFR/TH/91-40 (1991).}
\lref\md{M. Duff, talk at Trieste summer 1989, Nucl. Phys. B335 (1990) 610.}
\lref\muller{M. Muller, Nucl. Phys. B337 (1990) 37.}
\lref\cfmp{C. Callan, D. Friedan, E. Martinec and M. Perry,  Nucl.
Phys. B262 (1985) 593.}
\lref\kir{E.B. Kiritsis, Mod. Phys. Lett. A6 (1991) 2871.}
\lref\vero{M. Ro\v cek and E. Verlinde, IAS preprint IASSNS-HEP-91/68
(hepth@xxx/\-9110053).}
\lref\giva{P. Ginsparg and C. Vafa, Nucl. Phys. B289 (1987) 414 \semi
E. Alvarez and M. Osorio, Phys. Rev. D40 (1989) 1150.}
\lref\grvsw{A. Giveon, E. Rabinovici and G. Veneziano, Nucl. Phys. B322 (1989)
167\semi
A. Shapere and F. Wilczek, Nucl. Phys. B320 (1989) 669.}
\lref\gz{M.K. Gaillard and B. Zumino, Nucl. Phys. B193 (1981) 221.}
\lref\cfg{S. Cecotti, S. Ferrara and Girardello, Nucl. Phys. B308 (1988) 436.}
\lref\iltq{L.E. Ib\'a\~nez, D. L\"ust, F. Quevedo and S. Theisen
unpublished (1990) \semi G. Veneziano, Phys. Lett. B265 (1991) 287.}
\lref\bmq{C. Burgess, R. Myers and F. Quevedo unpublished (1991) \semi
A. Tseytlin, Mod. Phys. Lett. A6 (1991) 1721.}
\lref\bv{R. Brandenberger and C. Vafa, Nucl. Phys. B316 (1989) 391.}
\lref\mv{K. Meissner and G. Veneziano, Phys. Lett. 267B (1991) 33\semi
M. Gasperini and G. Veneziano, CERN-TH-6321/91 (hepth@xxx/\-9112044).}
\lref\tv{A. Tseytlin and C. Vafa, Harvard preprint HUTP-91/A049
(hepth@xxx/\-9109048).}
\lref\dvv{R. Dijgkraaf, E. Verlinde, and H. Verlinde, IAS preprint
IASSNS-HEP-91/22\semi
A. Giveon, Mod. Phys. Lett. A6 (1991) 2843.}
\lref\rAGG{L. Alvarez-Gaum\'e and P. Ginsparg, Ann. Phys. 161 (1985) 423.}
\lref\rpglh{P. Ginsparg, ``Applied conformal field theory,''
Les Houches lectures (summer, 1988), published in
{\it Fields, Strings, and Critical Phenomena\/},
ed.\ by E. Br\'ezin and J. Zinn-Justin, North Holland (1989).}
\lref\rvafa{C. Vafa, ``Topological Mirrors and Quantum Rings'',
Harvard preprint HUTP-91/A059 (hepth@xxx/\-9111017).}
\lref\px{P. Candelas, X.C. de la Ossa, P.S. Green and
L. Parkes, Nucl. Phys. B359 (1991) 21; Phys. Lett. 258B (1991) 118.}
\lref\barst{I. Bars, University of Southern California preprint USC-91/HEP-B4.}
\lref\hh{J.H. Horne and G. Horowitz,
University of California preprint UCSBTH-91-39 (hepth@xxx/9108001).}
\lref\hhs{J. Horne, G. Horowitz and A. Steif, University of California preprint
UCSBTH-91-53 (hepth@xxx/9110065).}
\lref\giro{A. Giveon and M. Ro\v cek, IAS preprint IASSNS-HEP-91/84
(hepth@xxx/\-9112070).}
\lref\ind {S.P. Khastgir and A. Kumar Bubaneswar, preprint
(hepth@xxx/9109026).}
\lref\cresc{M. Crescimanno, Berkeley preprint LBL-30947\semi
P. Ho\v rava Chicago preprint EFI-91-57 (hepth@xxx/9110067)\semi
I. Bars and K. Sfetsos, Univ. Southern California preprints
USC-91/HEP-B5 (hepth@xxx/\-9110054) and USC-91/HEP-B6 (hepth@xxx/\-9111040)
\semi D. Gershon, preprint TAUP-1937-91 (hepth@xxx/9202005).}
\lref\dlp{L. Dixon, J. Lykken and M. Peskin, Nucl. Phys. B325 (1989) 325.}
\lref\barsc{I. Bars, Nucl. Phys. B334 (1990) 125.}
\lref\hel{S. Helgason, ``Differential Geometry, Lie Groups, and Symmetric
spaces'', Academic Press (1978)\semi
R. Gilmore, ``Lie Groups, Lie Algebras and Some of Their Applications'', Wiley
(1974).}
\lref\host{G. Horowitz and A. Strominger, Nucl. Phys. B360 (1991) 197.}

\newsec{Introduction}

Issues of principle in quantum gravity are among the most important
questions that may eventually be addressed in the context of string
theory. The study of curved spacetimes as string backgrounds could be used
to investigate the string theoretical approach to physics at the Planck scale
where classical methods are known to fail. The singularities in black hole and
cosmological backgrounds are especially interesting.

The description of string backgrounds has been studied extensively by means of
conformal field theory (CFT), but most of the effort thus far has been
directed to the case where the noncompact part of the spacetime is flat,
i.e.\ described by a trivial CFT, and only the internal space requires
nontrivial CFT techniques. Coset models provide a rich class of
explicit CFT's for this case and lead to an understanding of the space of
static tree-level vacua. Noncompact coset models provide a natural framework
for nonstatic backgrounds and other nontrivial spacetimes which have recently
received more attention, especially in the context of 2D gravity.

In this article we expand on previous approaches \refs{\bn,\witten} to study
noncompact cosets in the Wess-Zumino-Witten (WZW) formalism \refs{\rwzw,\wzwg}.
We present a general discussion of such spaces,
classifying all those with a single timelike coordinate and any
number of spacelike coordinates. These will provide the natural background
for consistent string propagation. We also discuss
the structure of singularities that occur in gauged WZW models, as well as the
geometrical interpretation of the spaces obtained in this way.
The metric obtained from the WZW model is singular and we show that there
are singularities at fixed points of the gauge transformation. In cases
where there is a dual gauging, these same group elements will provide
horizons of the dual metric, generalizing the horizon/singularity duality.

We shall then discuss in detail a simple class of models that provides an
example for any spacetime dimension: $SL(2,\IR)\otimes
SO(1,1)^{D-2}/SO(1,1)$. For $D=2$, this is known
to describe a single self-dual black hole geometry. We find that for $D>2$,
there are two geometries described by the two anomaly-free gaugings
(vector and axial). Unlike the Lorentzian $D=2$ case in which both gaugings
give dual versions of the same
black hole, the two geometries in this case are seemingly different. They are
nonetheless dual to each other in a manner similar to
mirror manifolds in string compactifications, where a single CFT can
describe different target space geometries. For the axial gauging, we
find the geometry (2D black hole)$\otimes \IR^{D-2}$, or a $D-2$ black brane
\host.
For the vector gauging, on the other hand, we find (3D black string)$\otimes
\IR^{D-3}$ with nonvanishing torsion which is a different black brane.
We explicitly verify that the large $k$ (Kac-Moody level) limit of the gauged
WZW model gives a solution of the field
equations for dilaton, graviton and antisymmetric tensor field to lowest
order in $\alpha'$. Unlike the 2D case, however, we are able to find a more
general solution of those equations. The explicit
form of the curvature scalar is used to clarify the nature of the
singularities leading to the above black hole interpretation. Duality acts
in a nontrivial way for these geometries, in particular the
singularity for the axial gauging occurs for elements of the coset
that are not fixed points, and its dual is at a regular surface in the
asymptotically flat region of the vector gauged geometry. This
illustrates the possibility that strings do not necessarily preclude spacetime
singularities but may nonetheless be better behaved than expected on them.

In section 2 we present a general discussion of the geometry of noncompact
coset spaces and their associated Kac-Moody algebras, and their
description in terms of gauged WZW models. We discuss the conditions for
anomaly-free gaugings and the conditions for the existence of isometries.
The classification of all coset WZW models with a single timelike
coordinate is a purely group theoretical question which we solve using the
known properties of general coset spaces. Finally we discuss briefly the
appearance of singularities in the large $k$ limit of gauged WZW models
and their relation to fixed points of the corresponding gauge
transformation.

In section 3 we construct explicitly the metric for the models
$SL(2,\IR)\otimes SO(1,1)^{D-2}/SO(1,1)$ for the two anomaly-free gaugings
and obtain the geometries mentioned above. We explicitly verify that the
background fields obtained from the large $k$ limit of the WZW
model satisfy the string equations to lowest order in $\alpha'$. We also
find the most general solution of those equations with the same number of
isometries, and argue that it can be obtained from marginal perturbations
of the coset CFT. Finally in section 4 we discuss duality of these
solutions. We briefly review the duality of $\sigma$--models whenever there
is an isometry (following reference \busc),
and generalize to the case of several commuting isometries.
This symmetry in particular relates spaces with
torsion to spaces without torsion. We prove the relation between the
two different geometries by identifying the vector transformation
($g\to hgh\inv$) as the isometry when gauging the axial transformation
($g\to hgh$), and vice-versa. We end with some final comments and future
developments, and compare our work with other recent papers on the subject.

\newsec{Noncompact coset CFT's}

\def\tdbhtext{The causal structure of the two dimensional black hole spacetime
of \witten. Regions I,IV are asymptotic regions, regions II,III are inside
the horizon, and regions V,VI are beyond the singularities.}

The study of noncompact coset CFT's was undertaken in \dlp\
for SL(2,\IR)/$U(1)$ current algebra via the conventional
GKO construction. The formalism was later generalized to
any coset in \barsc. Given a level $k$
Kac-Moody algebra for a noncompact group $G$,
\eqn\sug{J_A(z)\, J_B(w)
\sim -{k\,\eta_{AB}/2\over (z-w)^2}+ {i\, f_{AB}{}^C\,J_C(w)
\over (z-w)}\, }
(where $g\ \eta^{AB}= f_{AC}{}^D\, f_{BD}{}^C$ is the Cartan metric and $g$ is
the Coxeter number of $G$),
the stress-energy tensor for a CFT based on $G$ is given by the Sugawara from
\eqn\set{T_G(z)={\eta^{AB}\ \colon J_A(z)\, J_B(z)\colon \over(-k+g)}\ .}
The corresponding central charge is $c_G=k\ {\rm dim}\ G/(k-g)$. For the
coset $G/H$ with stress-energy tensor $T_{G/H}=T_G- T_H$, the central
charge is $c_{G/H}=c_G-c_H$. The only changes from the compact case are
the sign of $k$ and of course the use of noncompact structure constants
$f_{AB}{}^C$. (The extension to supersymmetric coset models,
discussed in \barsc, will not be considered here.) The spectrum and the
corresponding elimination of negative norm states is not yet entirely
understood for these models, and more progress is needed before we can
properly treat the string vacua obtained from this approach.

In \witten, it was shown that the SL(2,\IR)/$SO(1,1)$ current algebra could be
interpreted as a two dimensional black hole spacetime. An implicit prescription
for assembling the holomorphic and anti-holomorphic representations was given
in terms of a gauged WZW model.
In \fig\ftdbh{\tdbhtext}, we reproduce the causal structure of this spacetime.
In this section we shall provide some background on this construction,
emphasizing the semi-classical limit in which various aspects of the geometry
can be visualized, and which provides a geometric interpretation for the
GKO current algebra construction. For an exact incorporation of all quantum
corrections, however, we would need to return to the full
conformal field theory / current algebra approach.

\figinsert\ftdbh{\epsfxsize2in$$\epsfbox{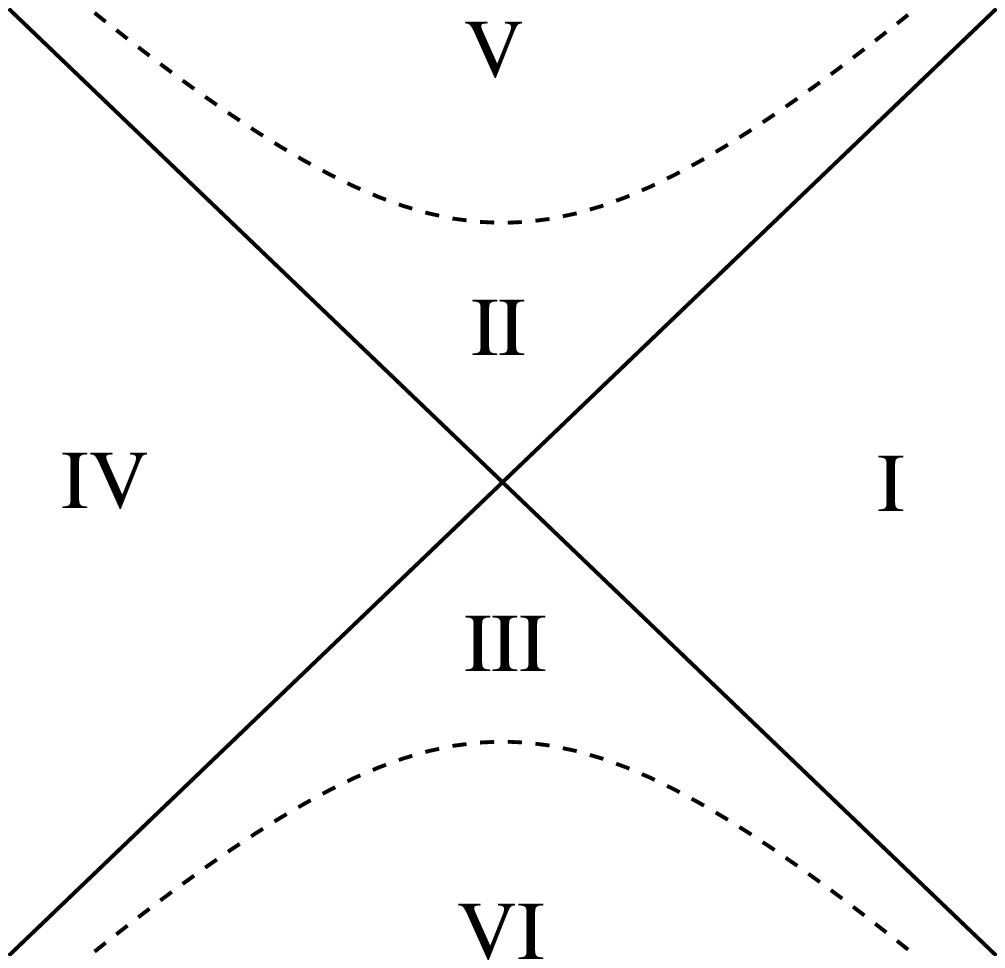}$$}{\tdbhtext}

\subsec{Anomalies}

The WZW action in complex coordinates is written
\eqn\wzcc{L(g)={k\over{4\pi}}\int \d^2z\, \tr(g\inv\del g\, g\inv\delbar g)-
{k\over{12\pi}}\int_{B}\tr (g\inv dg\wedge g\inv dg\wedge g\inv dg)\ ,}
where the boundary of $B$ is the 2D worldsheet.
To promote the global $g\to h_L\inv\, g\, h_R$ invariance to a local
$g\to h\inv_L(z)\,g\,h_R(\overline z)$ invariance, we let
$\del g\to \del g + Ag$, and $\delbar g\to \delbar g-g\bar A$. The gauge fields
transform as $A\to h_L\inv(A+\del)h_L$ and $\bar A\to h_R\inv(\bar
A+\delbar)h_R$ (so that $Dg\to h_L\inv Dg\, h_R$ for $D$ equal to either
holomorphic or anti-holomorphic covariant derivative).
Vector gauge transformations correspond to $h_L=h_R$, and axial gauge
transformations to $h_L=h_R\inv$.
Substituting covariant derivatives in \wzcc\ gives the gauged action
\eqn\wzwmcc{L(g,A)=L(g)+{k\over{2\pi}}\int \d^2z\,\tr\bigl( A\,\delbar g g\inv-
 \bar A \,g\inv\del g - g\inv A g \bar A \bigr)\ .}

Under the infinitesimal transformations $h_L\approx 1+\alpha$, $h_R=1+\beta$,
we have $\delta A=\del \alpha+[A,\alpha]$ and $\delta \bar A=\delbar
\beta+[\bar A,\beta]$. The anomalous variation of the effective action is (see
e.g.\ \rAGG\ for a review)
\eqn\evW{\delta W={k\over2\pi}(\alpha\delbar A+\beta\del\bar A)\ .}
The variation of the (LR$\to$VA) counterterm $\tr A \bar A$, on the other
hand, is
\eqn\eCT{\delta(\tr A \bar A)=\tr\Bigl(-\beta\delbar A-\alpha\del\bar A
+(\alpha-\beta)\bigl[\bar A,A\bigr]\Bigr)\ .}
For the abelian case, we see that \eCT\ can compensate the variation
\evW\ for either $\alpha=\pm \beta$ since the commutator term automatically
vanishes.
Thus both vector and axial-vector gauging are allowed. In the non-abelian case,
only the vector gauging $\alpha=\beta$ is allowed. (An integrated form of this
argument may be found in \kir: essentially the relevant $\pi_3$
obstruction is not captured by the trivial topology of $U(1)$.)
If we change sign $\bar A\to -\bar A$ for the axial gauged case (to give $A$
and $\bar A$ the same transformation properties), then the gauged action may
be written
\eqn\wzwcc{L(g,A)=L(g)+{k\over{2\pi}}\int \d^2z\,
 \tr\bigl( A\,\delbar g g\inv\mp\bar A \,g\inv
 \del g +A \bar A \mp g\inv A g \bar A \bigr)\ ,}
where the upper and lower signs represent respectively vector ($g\to hgh\inv$)
and axial-vector ($g\to hgh$) gauging.

It is intuitively reasonable that gauging $g\to hgh^{\mp1}$ should result
in a combination of holomorphic and antiholomorphic representations of
$G/H$ algebras: since the equations of motion for the ungauged model
result in $g(z,\overline z)= a(z)\,b(\overline z)$, gauging left
multiplication by $h(z)$ and right multiplication by $h^{\mp1}(\overline
z)$ properly removes the $H$ degrees of freedom from {\it both\/} sides.

\subsec{Semi-classical limit}

\def\rottext{Under rotation about the origin, any point in the plane can be
gauge-fixed to the positive real axis. The origin is left fixed.}
\def\trcigtext{The two dual versions of the Euclidean black hole. The upper
``trumpet'' version has a singularity at $r=0$, reflecting the fixed point at
the origin of \frot. The lower ``cigar'' version is free of singularities.}
\def\lctext{Under Lorentz boost, the plane is partitioned into four regions
bounded by the light cones. The points
in each region can be gauge-fixed to the indicated dotted lines. The origin is
left fixed.}
\def\hypertext{The SL(2,\IR) hyperboloid with the $x$ coordinate suppressed.
The two black dots represent the fixed points of \flc, and the gray lines
represent the gauge-fixing. The two lightcones are the intersections of the
hyperboloid with planes perpendicular to the $y$ axis at $y=\pm1$.}

We now consider some naive properties of the geometry described by \wzwcc\
in the large $k$ (semi-classical) limit.
Writing $A=A^a\sigma_a$ in terms of the generators $\sigma_a$ of $H$,
and integrating out the components $A^a$ classically gives the effective
action
\eqn\effcc{L=L(g)\pm{k\over{2\pi}}\int \d^2z\
 \tr(\s_b g\inv\del g)\, \tr(\s_a\delbar g g\inv)
 \,\Lambda_{ab}\inv\ ,}
with $\Lambda_{ab} \equiv \tr(\s_a \s_b \mp \s_a g \s_b g\inv)$. Notice that
singularities of $\Lambda$ occur at least at fixed points of the gauge
transformation $g\to h g h^{\mp1}$. This is because for infinitesimal
$h\approx 1+\alpha^a\,\sigma_a$, we see that a fixed point $g$ satisfies
$\sigma_a\,g\mp g \sigma_a=0$. Multiplying by $g\inv \sigma_b$ and taking the
trace, we see that $\Lambda=0$ at a fixed point. (In the euclidean case it is
easy to show the converse, i.e.\ that $\Lambda=0$ implies a fixed point,
whereas this is not true in the lorentzian case.)

{}From the transformation properties of the gauge fields and \eCT, we note
that in the case of $H$ abelian the ungauged axial or vector symmetry
remains a {\it global\/} symmetry, i.e.\ an isometry of the spacetime
geometry. In the non-abelian case, not even a global vestige of the
ungauged symmetry remains. In the abelian case, this implies that a fixed
point of the ungauged symmetry corresponds to a point with vanishing
Killing vector. For lorentzian signature, the surface carried into the fixed
point by the isometry will be a null surface (the norm of the Killing
vector is conserved), in general nonsingular and hence a horizon. We see that
fixed points of symmetry transformations generically give rise to metric
singularities when the symmetry is gauged and to horizons when ungauged. This
general property is the origin of the singularity/horizon duality of the
2D black hole of \witten. For the vector gauging, the metric can be written
$\d s^2=-\d a\,\d b/(1-ab)$, and the fixed point of the vector transformation
(the gauged symmetry) corresponds to $ab=1$, which is the singularity. The
fixed point of the axial transformation (the isometry) is $a=b=0$
indicating that the invariant surface $ab=0$ is null, and provides the
event horizon illustrated in \ftdbh. For the axial gauging, the
metric is identical (i.e.\ the geometry is self-dual) but the role of the
fixed points is exchanged,  implying the horizon/singularity duality
pointed out in \dvv.

We now try to visualize in more detail the naive properties of the
geometry described by \wzwcc\ for $G=\rm SL(2,\IR)$ in the large $k$
(semi-classical) limit. We take SL(2,\IR) group elements parametrized
as\foot{The compact group $SU(2)$ has group elements
$g=x_0\one + i\vec x\cdot\vec \sigma$ so we identify $\sigma_1$ and
$\sigma_3$ as the noncompact generators, giving signature (-- + --).}
\eqn\eslge{g=w\one + x\sigma_1 +iy\sigma_2 + z\sigma_3
=\pmatrix{w+z & x+y\cr x-y & w-z\cr}\ .}
The condition that $\det g=1$ requires $w^2+y^2-x^2-z^2=1$, so we see
that $x,z$ parametrize the $\IR^2$ and
$w,y$ the $S^1$ of the $\IR^2\times S^1$ topology of SL(2,\IR).
We shall consider the actions $g\to hgh^{\mp1}$ for $h=h_{E,L}$, where
\eqn\ehehl{h_E\equiv\ee{i \alpha \sigma_2}\ ,
\quad h_L\equiv\ee{\alpha \sigma_3}\ .}
Modding out by $h_E$ gives a Euclidean signature metric, and modding out by
$h_L$ gives a Lorentzian signature metric.

For the euclidean case $h=h_E$, the action $g\to h_E\,g\,h_E\inv$ is easily
seen to be rotation in the 1,3 = $x,z$ plane.
Thus we can always ``gauge-fix'' any point to, say, the positive $x$ axis
(\fig\frot{\rottext}). The origin
is a fixed point of the transformation and results in a singularity of the
metric. Crossing this half-line with endpoint singularity back together with
the $w,y$ circle, we see that the spacetime takes the form of a ``trumpet''
(\fig\ftrcig{\trcigtext}).

The other action $g\to h_E\,g\,h_E$ is simply rotation in the $w,y$ circle.
(This is easily seen by reparametrizing $g=\tilde g \sigma_1$, which exchanges
$(x,z)\leftrightarrow(w,y)$, and noting that
$g\to hgh\Rightarrow \tilde g\to h\tilde g h\inv$.)
By rotation of the $w,y$ circle, we can always ``gauge-fix'' every point to,
say, the point (1,0), thus eliminating the $S^1$ entirely. Since the action has
no fixed points the metric on the remaining $\IR^2$
can be entirely regular. For metrical reasons, we change to $r,\theta$
coordinates in which this $\IR^2$ is naturally depicted as a ``cigar''.
Fig.~\xfig\ftrcig\ thus depicts the two dual semiclassical geometries of the
Euclidean 2D black hole constructed in \witten.

\figinsert\frot{\epsfxsize1.75in\relax$$\epsfbox{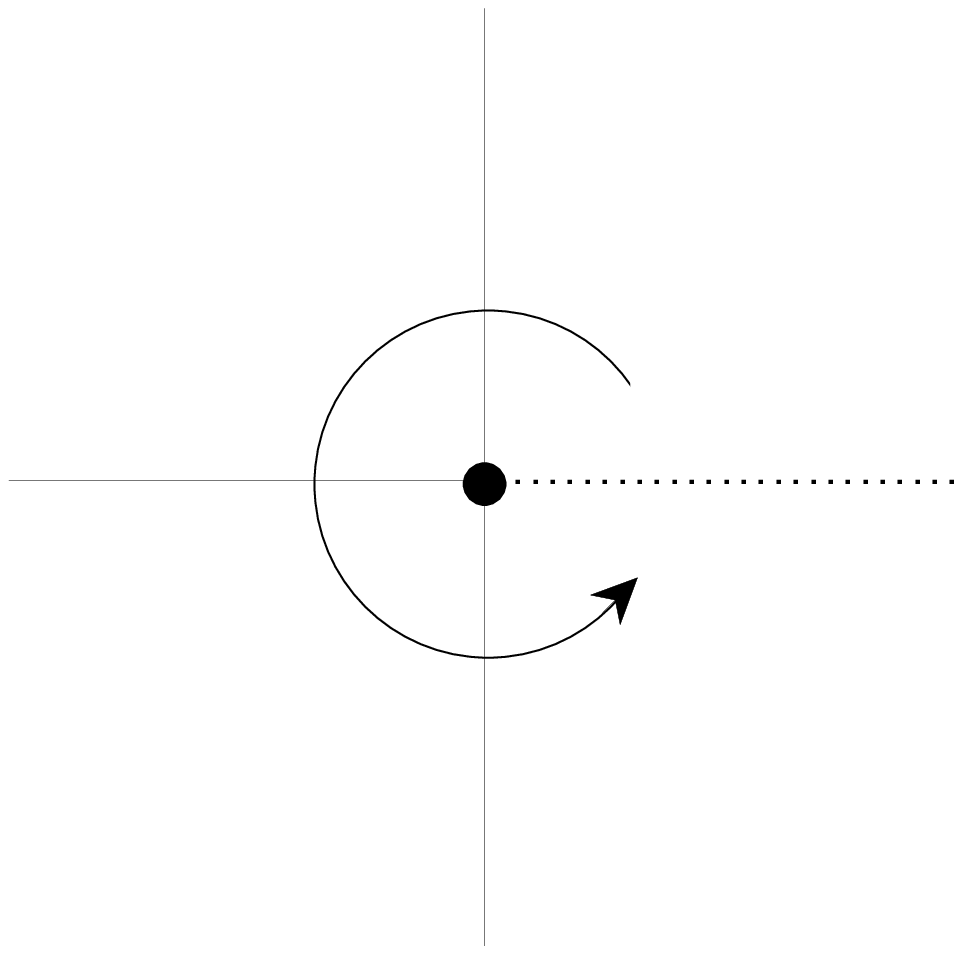}$$}{\rottext}
\figinsert\ftrcig{\epsfxsize2in$$\epsfbox{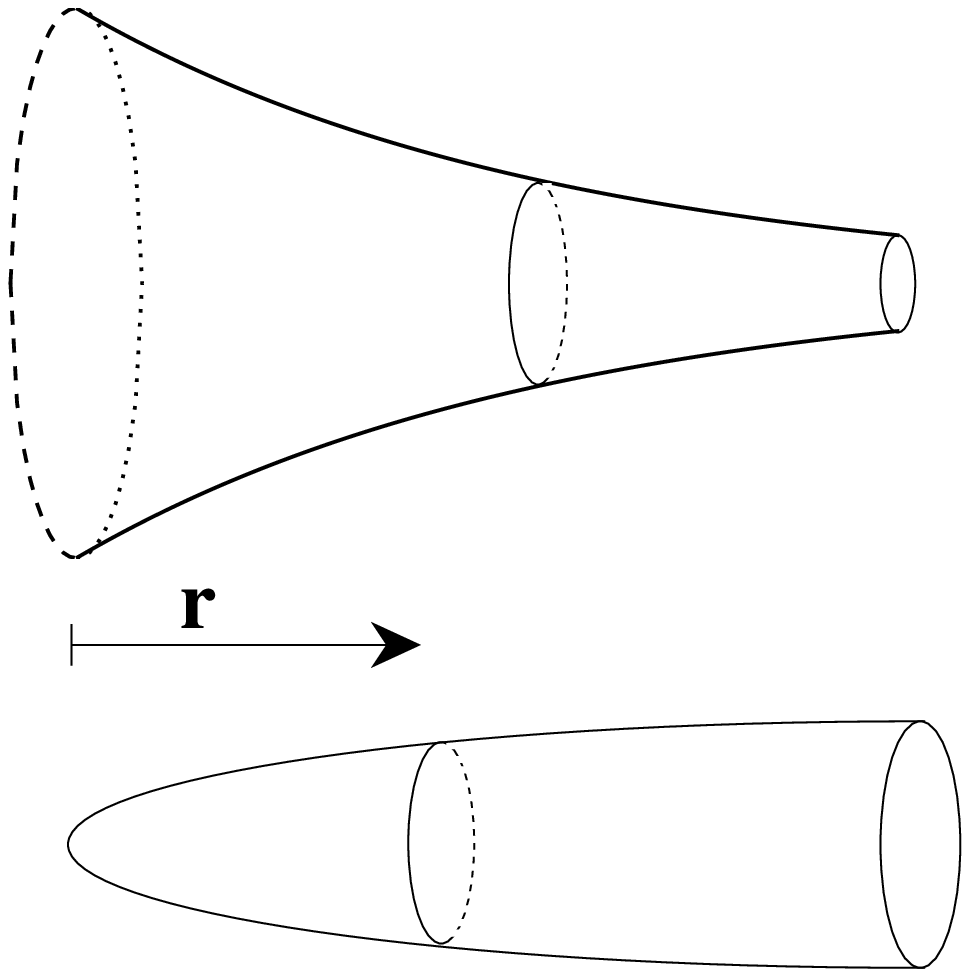}$$}{\trcigtext}

For the Lorentzian case $h=h_L$, on the other hand, the action $g\to h gh\inv$
is a boost in the $(x,y)$ coordinates,
and the action $g\to hgh$ is a boost in the $(w,z)$ coordinates. (The latter
result is easily seen by the same reparametrization $g=\tilde g \sigma_1$
mentioned above.) Up to trivial reparametrization, these two actions are
identical so we see that the Lorentzian version of the 2D black hole is
self-dual. Instead of the compact action of \frot\ and the gauge fixing to a
single ray, for the Lorentz boost we have the action depicted in
\fig\flc{\lctext}, which partitions the vicinity of the origin into four
disjoint regions and leaves the origin fixed.

\figinsert\flc{\epsfxsize1.75in$$\epsfbox{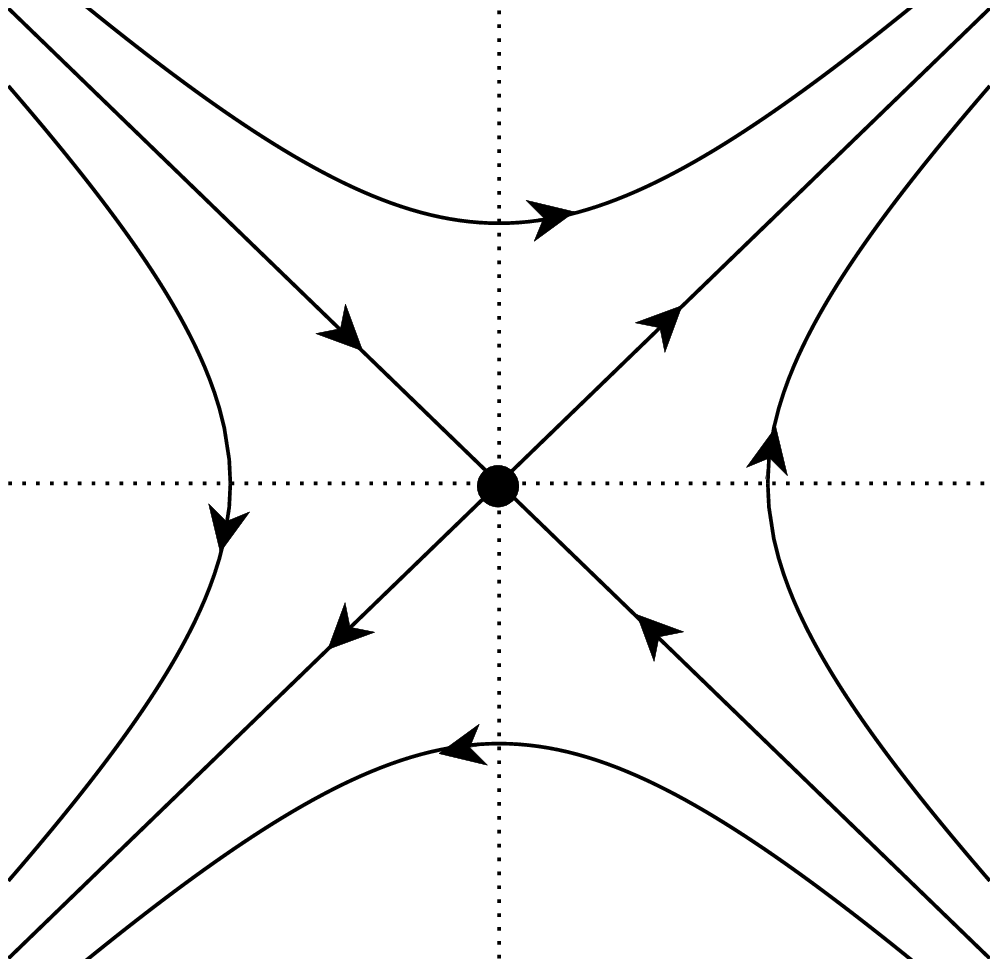}$$}{\lctext}

Modding out by the action $g\to hgh$, we
can gauge-fix the action of the $(w,z)$ boost to lines with $z=0$ and $w=0$
(which correspond respectively to $a=\pm b$ in the
parametrization $g={\ \,a\ \ u \choose -v\ \ b}$ used in \witten), as shown in
\flc. In \fig\fhyper{\hypertext}, we transcribe this picture to
the SL(2,\IR) hyperboloid (with the $x$ coordinate suppressed).
The region $z=0$ interpolates between the two fixed lines $\pm(i \sigma_2\cosh
\alpha + \sigma_1\sinh \alpha)$, passing along the $x$ direction through the
points $g=\pm i \sigma_2$ at $y=\pm1$ in \fhyper.
This $z=0$ region encompasses two copies of regions I--IV of \ftdbh. (The
origins of the lightcones of \ftdbh\ appear at the points $g=\pm\one$, i.e.\
$w=\pm1$, of \fhyper\ with the $x$ direction restored.)
The region $w=0$ encompasses two copies apiece of regions V and VI.
As mentioned earlier, the (self-) duality in this case corresponds to
interchanging $w\leftrightarrow y$, $z\leftrightarrow x$, so we can see from
\fhyper\ how the duality exchanges region I (which is the asymptotic region
moving upwards along the $x$ direction from $w=1$) with region V, and similarly
IV$\leftrightarrow$VI (see \dvv). Region II (which lies between $w=1$ and
$y=1$) is mapped into itself, as is region III.

\figinsert\fhyper{\epsfxsize3in$$\epsfbox{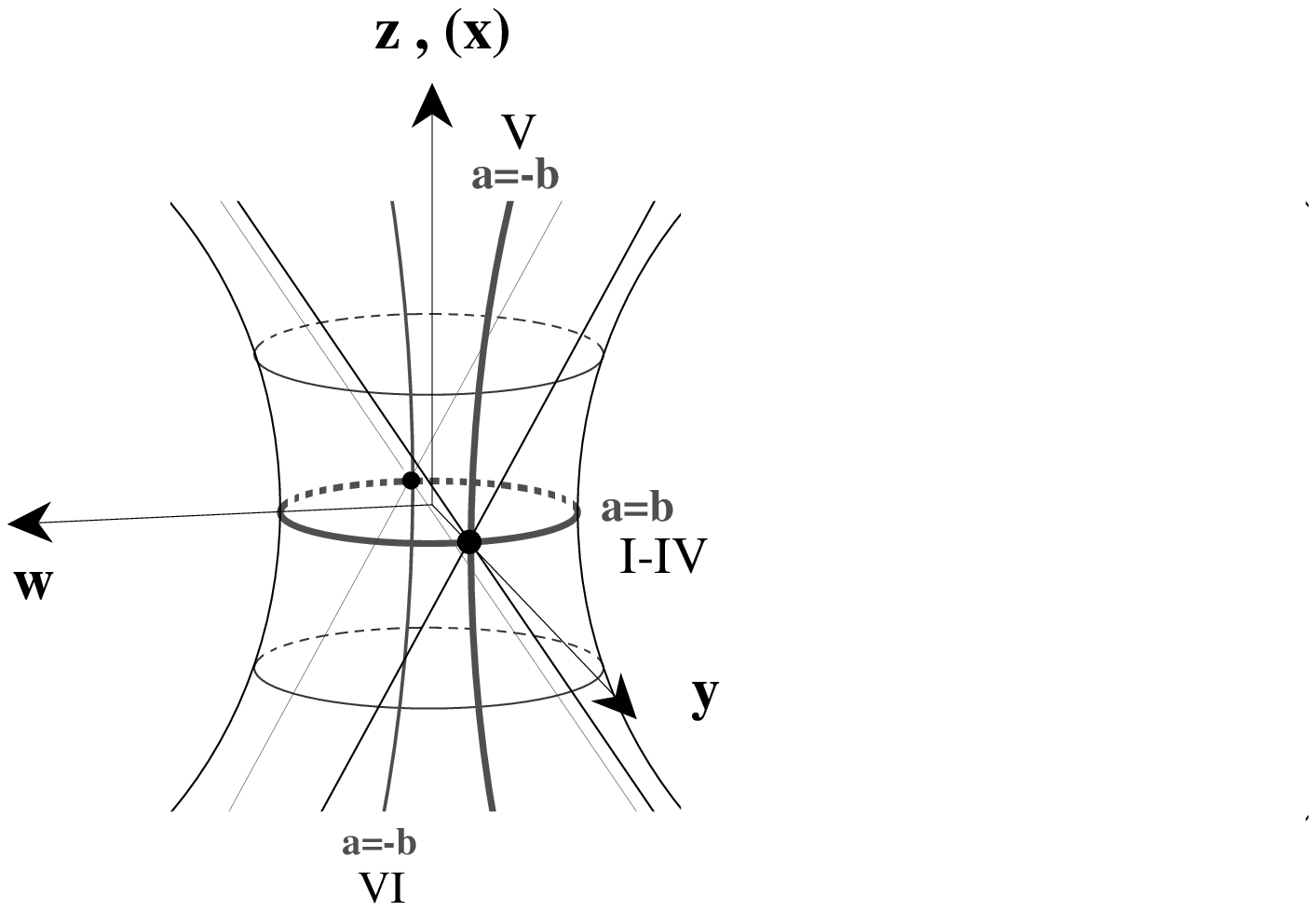}$$}{\hypertext}

(Finally, for comparison with the semiclassical limit of compact cosets, we
point out that the case of $SU(2)/U(1)$ gives a disk with a singular
boundary. Recall that $g\in SU(2)$ can be parametrized as
$g=\cos\chi + i\hat n\cdot \vec\sigma\,\sin\chi$, where $\chi\in [0,\pi]$
denotes an azimuthal angle on $S^3$ and $\hat n$ parametrizes latitudinal
$S^2$'s with radius $\sin\chi$. The action $g\to h g h\inv$ with
$h=\ee{i\alpha \sigma_3/2}$ simply rotates $\hat n$ by angle $\alpha$ about the
3-direction, i.e.\ for $\hat n=(\sin\theta\cos\phi,\sin\theta\sin\phi,
\cos\theta)$ the action is the translation $\phi\to\phi+\alpha$. Modding out by
this action simply removes the $\phi$ coordinate and squashes each latitudinal
$S^2$ parametrized by $\hat n$ to an interval $\theta\in[0,\pi]$ with size
still proportional to $\sin\chi$. The result is a disk whose boundary, given by
the circle $\ee{i\alpha \sigma_3/2}$, is a line fixed under the group action
and consequently a singularity of the induced metric on the disk.
By the argument used above in the noncompact case, modding out by the dual
action $g\to hgh$ results in an equivalent picture.
(Write $g=x_0\one+i\vec x\cdot\vec\sigma$ as $i\tilde g\sigma_1$, which
interchanges $(x_0,x_3)\leftrightarrow(x_1,x_2)$,
and has $\tilde g\to h\tilde g h\inv$.) The $U(1)$ gauged $SU(2)$ WZW model
thus gives another example of a theory that is self-dual in this sense.)

\subsec{Enumeration of possibilities}

We have seen that the gauged ``$G/H$'' WZW models considered here are not
the usual left or right $G/H$ coset spaces with standard coset metric, as
considered in the mathematical literature \hel\ and in standard treatments
of coset space nonlinear $\sigma$--models\vhol. This is because we gauge $g\to
hgh^{\mp1}$ type symmetries rather than $g\to gh$ or $g\to hg$, and as well
we include a Wess-Zumino term which can add a torsion piece to the metric.
Gauging the $H$ subgroup nonetheless eliminates the $H$ degrees of
freedom, and it is easily verified that the {\it signature\/} of the
resulting metric is the same as that of the standard coset metric. It is
therefore straightforward to impose the phenomenological restriction to
spaces with only a single timelike coordinate \bn. The only subtlety is
that the level $k$ appears in front of the action. Positive $k$,
in our sign conventions, results in a metric whose
compact generators correspond to timelike directions and noncompact generators
to spacelike directions. For
negative $k$ (when allowed by unitarity), the roles of
compact and noncompact generators are interchanged in the correspondence.

To classify all the coset CFT's with a single timelike coordinate we consider
first the case $k$ positive and examine the difference
$$N\equiv |G|_{\rm c} - |H|_{\rm c}\ ,$$
for all possible cosets, where $|G,H|_{\rm c}$ denote the number of compact
generators. To this end we employ the known classification \hel\ of
symmetric spaces $G/H$ (where $H$ is a maximal subgroup and $G$ is simple).
{}From this list we eliminate all cases
with $N>1$, since for a given $G$ modding out by smaller (non-maximal)
subgroups increases the value of $N$. For $N=1$, this leaves only
the case $SO(D-1,2)/SO(D-1,1)$ (\bn). For $N=0$, which corresponds to
maximal compact subgroup embeddings, the possibilities are listed in table 1.
{}From this table we identify the cases for which $H$ has a $U(1)$ factor,
$H=H'\times U(1)$ (hermitian symmetric spaces), so that $G/H'$ has an
additional compact generator, hence one timelike coordinate.
These latter cases are listed in table 2, and
exhaust all possibilities in which $G$ is a simple group.
For $k$ negative, we consider instead the difference
$N=|G|_{\rm nc} - |H|_{\rm nc}$ of noncompact generators, and find that the
only solution with $N=1$ is $SO(D,1)/SO(D-1,1)$.

For $G$ a product of simple groups and $U(1)$ factors, there are several
possibilities to consider:

\item{(i)\ } $G=G_1\otimes G_2\otimes G_3$ and $H=H_1\otimes H_2 \otimes H_3$
where $G_1/H_1$ is in table 2, $G_2/H_2$ is in table 1 (or products thereof)
and $G_3/H_3$ is a (product of) compact coset(s).

\item{(ii)\ } $G=G'\otimes \IR$ where $G'/H$ has $N=0$ (products of cases from
table 1 and compact). In this case $\IR$ provides the timelike coordinate.

\noindent
These are the most general cases. Possibilities such as products of cases
in table 2 modded out by several U(1)'s, for example, are already included
in case (i). In table 3, we list all such cases with coset dimension $\leq
10$ (and due to space limitations omit those with $G$ compact). In this
table, it is implicitly understood that all possible embeddings $H\subset
G$ are to be considered. The number of possible models so obtained is
relatively small, particularly for lower dimensions.

Other possibilities may be obtained by enlarging consideration from
semisimple groups $G$ to non-semisimple groups of potential relevance,
including of course the Poincar\'e group. In 3D, for example, ignoring the
non-semisimple cases leaves only $U(1)^3$, SL(2,\IR) and $SU(2)$, whereas
including them gives the nine groups corresponding to the Bianchi models
considered in cosmology. We discuss briefly how to treat other potentially
interesting cases involving non-semisimple groups.\foot{We thank V. Kaplunovsky
for a question that prompted this discussion.}
Let $G$ be non-semisimple, then it is a semidirect product of $S$
(a semisimple piece) and $R$ (the radical, or maximal invariant subgroup) \hel.
The algebra takes the form
\eqn\nss{[S,S]=S\ ,\qquad [S,R]\in R\ , \qquad [R,R]=R'\ ,}
where $R'\subset N \subset R$.
The Cartan matrix has zero eigenvalues, so
the group manifold itself is not interesting, but
modding out by subgroups may eliminate these zeroes to give a sensible
space with a well-defined metric. The number of zeroes is the
dimension of $N$, the maximal nilpotent subalgebra of $R$.
The possible cosets are

\item{(1)\ } $G/N$. Since $G/N=S\otimes R/N$
    and $R/N$ is an abelian invariant subalgebra, the zeroes are not eliminated
    but instead are equal in number to the dimension of $R/N$.

\item{(2)\ } $G/R=S$. This case gives all the semisimple
    groups. Those with a single timelike coordinate are $SL(2,R)\otimes C$ (we
    could also include $C\otimes SO(1,1)^n$),
    where $C$ is any compact semisimple group and one of the
    $SO(1,1)$'s has negative level $k$ to provide the timelike direction.

\item{(3)\ } $G/S=R$. In general $R$ is an invariant
    subalgebra and will have abelian subalgebras with associated zero
    eigenvalues, so the zeroes are not eliminated. The only exception is when
    $R$ itself is abelian, so the Cartan
    metric is not defined from the regular representation and will
    be nonsingular. This leads to interesting cases such as
    $ISO(d-1,1)/SO(d-1,1)$, but the general classification is not known.
    Whenever $R$ as a group has only a single timelike coordinate, the zeroes
    can be eliminated but since $R$ is abelian the only choices are products
    of $SO(1,1)$'s and $U(1)$'s.

\item{(4)\ } $(G/S)/N=R/N$. As mentioned above this is an abelian subalgebra,
    so the Cartan metric is not defined from the regular representation
    and we are left with the situation of (3).

\item{(5)\ } A more general situation would be to mod out by different
    subgroups not in the above decomposition, but this probably gives nothing
    new since $R$ is a semidirect
    product of abelian groups whose survival in the coset
    $G/H$ (for any $H$) would result in zero eigenvalues of the metric unless
    everything remaining is abelian as in (3) and (4) above. If they are all
    eliminated then we revert to case (2) above.

\nobreak\noindent
A general means of obtaining non-semisimple cosets is by group contractions,
so it may well be possible to find a more systematic and complete procedure
to generate all non-semisimple cosets using the properties of the semisimple
cases.

\newsec{$SL(2,\IR)\otimes SO(1,1)^{D-2}\big/SO(1,1)$ Models }

We now consider the simplest class of coset models with a single
timelike coordinate and any number of spacelike coordinates. In order to
find the metric in the large $k$ limit, we employ the standard procedure
in nonlinear $\sigma$--models \vhol, as outlined in section 2:
i.e.\ find a parametrization of
the $G$ group elements, impose a unitary type gauge on the fields in the
$\sigma$--model action and then solve for the (non-propagating) $H$-gauge
fields to derive the $G/H$ worldsheet action. From that action
we can read of the corresponding background fields. For the sake of
generality, we write down the integrated action for a generic, not
necessarily simple, group $G$ and a subgroup $H$, not necessarily abelian.

We first write the gauged WZW action \wzcc,\wzwcc\ as
\eqn\wzw{L(g,A)=L(g)+{1\over{2\pi}}\sum_i k_i\int \d^2z\,
 \tr\bigl( A\,\delbar g g\inv\mp\bar A \,g\inv
 \del g +A \bar A \mp g\inv A g \bar A \bigr)_i\ ,}
where $\mp$ represents respectively vector ($g\to hgh\inv$) and axial-vector
($g\to hgh$) gauging.
The ungauged action is
\eqn\wz{L(g)={1\over{4\pi}}\sum_i k_i\int \d^2z\, \tr(g\inv
  \del g\, g\inv\delbar g)_i-
  {1\over{12\pi}}\sum_i\int_{B}\tr (
   g\inv dg\wedge g\inv dg\wedge g\inv dg)_i\ ,}
where $i$ runs over the simple group factors in $G=\otimes_i G_i$.
Writing $A=A^a\sigma_a$ in terms of the generators $\sigma_a$ of $H$,
and integrating out the components $A^a$ classically gives the effective
action
\eqn\eff{L=L(g)\pm{1\over{2\pi}}\sum_{i,j} k_i k_j \int \d^2z\
 \tr(\s_b g\inv\del g)_i\, \tr(\s_a\delbar g g\inv)_j
 \,\Lambda_{ab}\inv\ ,}
with
\eqn\matrl{\Lambda_{ab} \equiv \sum_l k_l\, \tr(\s_a \s_b \mp \s_a g \s_b
g\inv)\ .}
Notice that at the singular points of $\Lambda$ the classical integration
of the gauge fields fails, and hence
(according to the discussion following \effcc)
where singularities of the target space metric are expected.

For the $SL(2,\IR)\otimes SO(1,1)^{D-2}\big/SO(1,1)$ models, we
parametrize the group elements as
\eqn\ggg{g=\pmatrix{g\dup_0&0&\ldots&0\cr
    0&g\dup_1&\ldots&0\cr
    \vdots&\vdots&\ddots&\vdots\cr
    0&0&\ldots&g\dup_{D-2}\cr}\ ,}
where
\eqn\gz{g\dup_0=\pmatrix{a&u\cr -v&b\cr} \qquad({\rm with}\ ab+uv=1)}
and
\eqn\ggi{g\dup_i=\pmatrix{\cosh r_i&\sinh r_i\cr \sinh r_i&\cosh r_i\cr}
 \quad{\rm with}\quad i=1, ..., D-2 \ .}
We choose the embedding such that the generator of $H=SO(1,1)$ is
\eqn\gen{\s=\pmatrix{s_0&0&\ldots&0\cr
0&s_1&\ldots&0\cr
\vdots&\vdots&\ddots&\vdots\cr
  0&0&\ldots&s_{D-2}\cr}\ ,}
where
\eqn\gend{s_0=q_0\pmatrix{1&0\cr 0&-1\cr}\quad{\rm and}\quad s_i=q_i
  \pmatrix{0&1\cr 1&0\cr}\ ,}
with coefficients normalized to $\sum_{i=0}^{D-2} q_i ^2=1$.

\subsec{Vector gauging}

Under the infinitesimal vector gauge transformations
$\delta g=\varepsilon(\s g-g \s)$, the parameters transform as $\delta
a=\delta b=\delta r_i=0$, and $\delta u=2\varepsilon q_0 u$, $\delta
v=-2\varepsilon q_0 v$. The choices $u=\pm v$ thus fix the gauge
completely. From $\pm u^2=1-ab$, we are left with the parameters $a, b,
r_i$ as the $D$ spacetime coordinates. Substituting \ggg\ and \gen\ into
\eff, we find the action (for both gauge choices $u=\pm v$):
\eqn\effd{\eqalign{L={k_0\over{2\pi}}\int \d^2z\,
 \biggl(-{\del a\delbar b +\del b\delbar a\over 2(1-ab)}
 +\sum_i \kappa_i\Bigl(\delta_{ij}+{\kappa_j\eta_i
 \eta_j\over 1-ab}\Bigr)\del r_i\delbar r_j\cr
   +{\kappa_i\e_i\over 2(1-ab)}\Bigl((b\del a-a\del b)\delbar r_i
 +\del r_i (b\delbar a-a\delbar b)\Bigr)\biggr)\cr}}
where $\k_i\equiv k_i/k_0$ and $\e_i \equiv q_i/q_0$. (From \gend\ we see that
the $\eta_i$'s parametrize the embedding of $SO(1,1)$ into the factored
$SO(1,1)$'s in $G$.)

This action can be identified with a $\sigma$--model action of the form
\eqn\efft{S=\int \d^2z\,\bigl(G_{MN}+B_{MN}\bigr)\,\del X^M \delbar X^N}
to read off the background metric and antisymmetric tensor field (torsion).
We see that \effd\ gives for $D=2$ the (dual) black hole metric of
\witten\ $\bigl(\d s^2={-\d a\,\d b/(1-ab)}\bigr)$.
For $\k_i\to 0$, it reduces as expected to the 2D black hole and for
$\e_i \to 0$ gives the 2D black hole times $D-2$ flat coordinates, again
as expected since in this limit $H=SO(1,1)$ is completely embedded in
$SL(2,\IR)$. Note that for any $D$ there is no torsion, in particular the WZ
term can be seen to be a total derivative for our choice of gauge.
Furthermore we can observe that there are at least $D-2$ isometries since
the metric does not depend explicitly on the coordinates $r_i$. Finally,
note that the metric blows up only at the fixed point $ab=1$ which is the
point where \matrl\ vanishes and the classical integration is not
justified. The fixed point of the isometry $g\to hgh$ is at $ab=0$, which
we expect to lead to a horizon.

To further analyze this metric, we change to coordinates in
which it is diagonal (such coordinates are to be expected due to the large
number of isometries). We consider (as in the 2D case) the regions
bounded by the horizon and singularity (\ftdbh):
\eqn\eregs{(i)\  0<ab<1\,,\quad (ii)\  ab<0\,,
\quad (iii)\ ab>1\ .}
$(i)$ corresponds to the interior regions II, III;
$(ii)$ to the asymptotic regions I, IV ;
and $(iii)$ to the additional regions V,VI.

In the interior regions $(i)$,
we can change to coordinates  $t$, $X_0$, $X_i$ by defining
\eqn\chvi{\eqalign{a&=\sin t\ \ee{(X_0+m X_{D-2})}\cr
    b&=\sin t\ \ee{-(X_0+mX_{D-2})}\cr
    r_i&=N_{ij}\, X_j\ ,\cr}}
with
\eqn\numu{N_{ij}=\cases{
-{\textstyle\rho_j\over \textstyle\rho_i\sqrt {\kappa_i}}&$i=j+1$\cr
\noalign{\vskip2pt}
{\textstyle\sqrt{\kappa_{j+1}}\,\eta_i\, \eta_{j+1}
\over\textstyle \rho_{j+1}\rho_j}&$j\geq i$\cr
\noalign{\vskip2pt}
{\textstyle\eta_i\over \textstyle\rho_j(\rho_j^2+1)^{1/2}}&$i\leq j=D-2$\cr
\noalign{\vskip2pt}
\qquad 0&otherwise\ ,\cr}}
\eqn\mumu{m=-q_0 \rho^2\ ,}
and
\eqn\rou{\rho_l^2 \equiv \sum_{i=1}^l \kappa_i\,\eta_i^2\ ,
\quad{\rm and}\quad\rho\equiv \rho_{D-2}\ .}
The matrix  elements $N_{ij}$ satisfy the relations
\eqn\ndnd{\eqalign{
\sum_l \kappa_l N_{li} N_{lj} &= \delta_{ij}\qquad i,j\neq D-2\ ,\cr
\sum_l\kappa_l\, N_{l D-2}^2&=1/(\rho^2+1)\ ,\cr
{\rm and}\ \sum_l\kappa_l\,\eta_l N_{lj}&=0 \ {\rm for}\ j\neq D-2 \ .}}
In these coordinates the metric takes the diagonal form
\eqn\metru{\d s^2=
{k_0\over {2\pi}}\Bigl(-\d t^2+\tan^2 t\, \d X_0^2+
    \sum_{i=1}^{D-2} \d X_i^2\Bigr)\ .}

The remaining regions are described similarly. For the asymptotic regions
$(ii)$, we use
\eqn\chvii{\eqalign{a&=\sinh R\, \ee{X_0+mX_{D-2}}\cr
b&=-\sinh R\, \ee{-(X_0+mX_{D-2})}\cr
r_i&=N_{ij}\, X_j\ ,}}
with the same $m$ and $N_{ij}$ as above. In these coordinates
the metric takes the form
\eqn\metrd{\d s^2
={k_0\over {2\pi}}\Bigl(\d R^2-\tanh^{2}R\, \d X_0^2
+\sum_{i=1}^{D-2} \d X_i^2\Bigr)\ .}
Finally, in the regions $(iii)$ beyond the
singularity the new variables are defined by
\eqn\chviii{\eqalign{a&=\cosh R \ee{(X_0+mX_{D-2})}\cr
 b&=\cosh R \ee{-(X_0+mX_{D-2})}\cr
 r_i&=N_{ij} X_j}}
with metric
\eqn\metrt{\d s^2=
{k_0\over {2\pi}}\Bigl(\d R^2-\coth^{2}R\, \d X_0^2+
    \sum_{i=1}^{D-2} \d X_i^2\Bigr)\ .}

Using the symmetry $a\to -a$, $b\to -b$, we identify the geometry (2D
black hole)$\otimes \IR^{D-2}$. In particular the isometry generated by
$g\to hgh$ is now explicit (it is a linear combination of translation in
$X_0$ and the $X_i$'s).   
We can see how the associated Killing 
vector changes signature on each boundary: it is timelike in
\metrd\ and \metrt\ and spacelike in the region \metru\ in between.
In \effd, this was not explicit in the $a, b, r_i$ coordinates. Although we
have chosen
a general embedding of $H=SO(1,1)$ in all of $G$, the resulting
geometry nonetheless coincides with the
case $\eta_i=0$, where $SO(1,1)$ was embedded only in $SL(2,\IR)$. This is as
expected since the $SO(1,1)$ factors in $G$ are
are abelian and therefore transform trivially under $g\to hgh\inv$.
The spacetime diagram for the relevant 2D geometry was shown in \ftdbh.

\subsec{Axial gauging}

We now consider the axial gauging for which things are less trivial.
Under the infinitesimal gauge transformation
$\delta g=\varepsilon (\sigma g+g\sigma)$,
we see that $\delta u =\delta v =0$ and $\delta a=2\varepsilon q_0 a$,
$\delta b=-2\varepsilon q_0 b$, $\delta r_i=2\varepsilon q_i$.
A simple choice that fixes the gauge completely is $a=\pm b$. Using $\pm
a^2=1-uv$ leaves $u$, $v$, and $r_i$ as the spacetime coordinates.
The gauged WZW action for the axial gauging is \wzw\ with the lower (+) signs,
%
%
and integration over the gauge fields gives again \eff\ but with $\Lambda$
defined by the lower (+) sign in \matrl.
%
%
%
%
Substituting \ggg\  and \gen\ into \eff, and using the above gauge fixing
gives the effective action
\eqn\effs{\eqalign{L={k_0\over{2\pi}}\int
&\d^2z\,\biggl( \Bigl(\kappa_i\,\delta_{ij}
-{\kappa_i\,\kappa_j\,\eta_i\,\eta_j\over{1-uv+\rho}}\Bigr)
\del r_i\delbar r_j + {(u\del v-v\del u)
(u\delbar v-v\delbar u)\over 4(1-uv+\rho)}\cr
&\qquad\qquad -\half(\del u\delbar v + \del v\delbar u)
-{(u\del v+v\del u) (u\delbar v+v \delbar u)\over 4(1-uv)}\cr
&\qquad-{\kappa_i\,\eta_i\over 2(1-uv+\rho)}
\Bigl[(u\del v-v\del u)\delbar r_i - \del r_i (u\delbar v-v\delbar u)\Bigr]
\biggr)\ .}}

{}From this expression we can make the following observations. First,
unlike the vector gauging, there is nonvanishing torsion given by the term
in square brackets in \effs, even though the WZ term vanishes for the
gauge choice made. We also can see that the metric has singularities at
$uv=1$, which in 2D is the fixed point of the axial transformation, and also at
$uv=1+\rho$, which is not a fixed point. Again the lines $uv=0$
represent horizons, and the metric and torsion do not depend
on the $r_i$ variables so there are also the $D-2$ isometries $r_i\to r_i
+ {\rm constant}$. As in the vector case, the $D=2$ ($\kappa_i = 0$) limit
reproduces the 2D black hole of \witten. Furthermore the $\eta_i=0$
limit gives the geometry (2D black hole)$\otimes \IR^{D-2}$
(with vanishing torsion) as in the
vector case, recovering the self-duality of those solutions.

\def\tdbstext{A two dimensional slice of the three dimensional black string
metric \effs. In addition to the regions of \ftdbh, the regions VII,VIII
lie between the singularities and inner horizons.}

The general case is more conveniently studied via
variables that diagonalize the metric in different regions. We will
consider the analog of the three regions $(i),(ii),(iii)$
of the vector case \eregs, but with $a,b$ exchanged for $u,v$.
In principle we could add an additional region due to the extra metric
singularity which forms the inner horizon (\fig\ftdbs{\tdbstext}), but we shall
find it is already included as part of region $(iii)$.

\figinsert\ftdbs{\epsfxsize2in$$\epsfbox{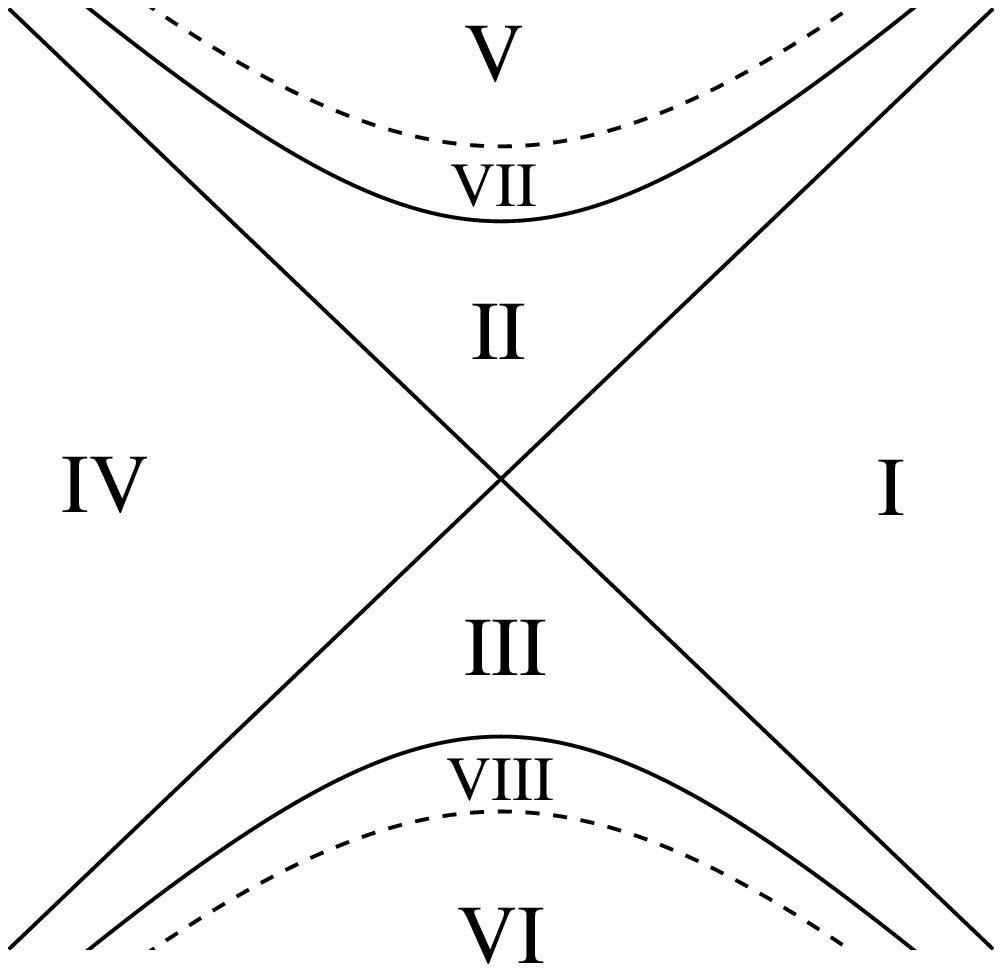}$$}{\tdbstext}

It is straightforward to see that the same changes of variables
\chvi,\chvii,\chviii\
made for the vector case also diagonalize this metric, but now for $m=0$
instead of \mumu.
For $(i)$ $0<uv<1$, we find
\eqn\metruu{\d s^2=
{k_0\over{2\pi}}\Bigl(-\d t^2+{1\over{(\rho^2+1) \tan^2 t +\rho^2}}
\bigl(\d X_0^2+\tan^2t\, \d X_{D-2}^2\bigr)+\sum_{l=1}^{D-3} \d X_l^2\Bigr)\ ,}
and the antisymmetric tensor is
\eqn\torsu{B_{X_0 X_{D-2}}=\bigl((\rho^2+1)\tan^2 t +\rho^2\bigr)\inv\ .}
In the region $(ii)$ $uv<0$, we have
\eqn\metrdd{\d s^2
={k_0\over{2\pi}}\Bigl(\d R^2+{1\over{(\rho^2 +1) \coth^2R-\rho^2}}
(-\d X_0^2 + \coth^2R\, \d X_{D-2}^2) +\sum_{l=1}^{D-3} \d X_l^2\Bigr)\ ,}
with torsion
\eqn\torsd{B_{X_0 X_{D-2}}=\bigl((\rho^2+1) \coth^2R-\rho^2\bigr)\inv\ .}
Finally in the region $(iii)$ $uv>1$ the metric is
\eqn\metrtt{\d s^2=
{k_0\over{2\pi}}\Bigl(\d R^2+{1\over{(\rho^2+1)\tanh^2R -\rho^2}}
(-\d X_0^2+\tanh^2R\, \d X_{D-2}^2)+\sum_{l=1}^{D-3} \d X_l^2\Bigr)\ ,}
with torsion
\eqn\torst{B_{X_0 X_{D-2}}=\bigl((\rho^2+1)\tanh^2R -\rho^2\bigr)\inv\ .}

{}From these metrics we can compute the corresponding curvature scalar
in each of the regions and find
\eqn\curv{R=B_{X_0 X_{D-2}} + {\rm constant}\ .}
We see that $R$ blows up only in region
$uv>1$ at the hyperbola $uv=1+\rho^2$ which is the real singularity,
whereas the surface $uv=1$ is only a metric singularity where the
signature of the metric changes. The latter is another horizon,  in addition to
$uv=0$. The geometry is thus (3D black string)$\otimes \IR^{D-3}$
with nonvanishing torsion and an inner horizon. The 2D representation
($uv$ diagram) with the eight different regions separated
by the horizons and singularity, is presented
in \ftdbs.
It is not surprising that there is a trivial $\IR^{D-3}$
crossed on since the $hgh$ action of $SO(1,1)$ only acts on one nontrivial
linear combination of $SO(1,1)$ generators of $G$.

We have thus far given the expressions for the metric and antisymmetric
tensor field for both gaugings, but not the expression for the dilaton.
This can be found in principle by considering the correct measure in the
path integral, but it is technically simpler to find it by solving the
background field equations to lowest order in $\alpha'$. This procedure
will also be useful to verify that the expressions we have given are
solutions of those equations. This is to be expected since they are valid
for large values of $k_0$ (equivalent to $1/\alpha'$ in the
sigma model expansion).

Let us consider the string background equations \cfmp
\eqn\einst{R_{MN}+D_M D_N \Phi -{1\over 4}H_M^{LP} H_{NLP}=0}
\eqn\hmn{D_L H^L_{MN}-(D_L\Phi)H^L_{MN}=0}
\eqn\dil{R-2\Lambda-(D\Phi )^2+2D_M D^M\Phi-{1\over 12}H_{MNP}H^{MNP}=0\ ,}
where $\Lambda\equiv (D-26)/3$ is the cosmological constant in the
effective string action and as usual $H_{MNP}\equiv \del_{[M} B_{NP]}$. To
check whether the expresions obtained above for the metric and antisymmetric
fields satisfy these equations, we can restrict to one of the regions.
We choose the ``cosmological'' region $0<uv,ab<1$ and assume an ansatz
\eqn\ans{\d s^2=-\d t^2+\sum_{i=1}^{D-1}r_i^2(t)\, \d X_i^2\ ,}
$\Phi=\Phi(t)$, and $H_{MNP}=H_{MNP}(t)$.
The two cases above are particular cases of this ansatz.
For the  vector gauging, $r_i=\rm constant$ for $i\geq 2$ and $H_{MNP}=0$.
For the axial gauging $r_i=\rm constant$ for $i\geq 3$ and $H_{MNP}$ is
nonvanishing only for $M,N,P=0,1,2$. The case without torsion was
solved in general in \muller\ and has solutions
\eqn\mull{r_i (t)=\alpha_i\tan^{p_i}\gamma t\ ,\qquad \sum_i p_i^2=1}
\eqn\mulld{\ee\Phi =\beta \tan^{2p} \gamma t \sec^2 \gamma t}
with $2p=1+\sum p_i$,  $\alpha_i$ and $\beta$
arbitrary constants and $\gamma^2\equiv {(26-D)/6}$.
It is easy to see that our solution for the vector gauging is a
particular case of this class of solutions with $p_i=0$, $i>1$, as long
as we make the shift $k_0\to k_0-2$. This is suggested by the relation
\eqn\cc{c={3 k_0\over{k_0-2}}-1+(D-2)=26}
which implies $(k_0-2)\inv={{26-D}/6}=\gamma^2$).
This representation makes it straightforward to discuss the
limit $D\to 26$ ($\gamma\to 0$) of our
solutions. Expanding $\tan \gamma t$ and rescaling the variables
we see that the metric depends on powers of $t$ \muller\
in the cosmological region and similarly for the other regions.
The curvature scalar behaves similarly so there is no singularity and
the black hole picture disappears. For $D>26$ ($k_0>2$), if
allowed by unitarity constraints (which have not yet been entirely clarified
for noncompact cosets)m we can analytically continue the
coordinates and get back the same black hole picture as
for $D<26$ with the interchanges II$\leftrightarrow$I(V)
and III$\leftrightarrow$IV(VI).

For the axial case the ansatz \ans\ substituted into \einst--\dil\ gives
the equations
\eqn\einstz{-\sum_i {\ddot{r_i}\over{r_i}}+\ddot\Phi-
\sum_{i<j}{\dot{B}_{ij}^2\over{2r_i^2 r_j^2}}=0}
\eqn\einsto{{\ddot{r_i}\over{r_i}}+\sum_{j\neq i}
 {\dot{r}_i\dot{r}_j\over{r_i r_j}}-
 {\dot r_i\over r_i}\dot\Phi+
 \sum_{j\neq i}{\dot{B}_{ij}^2\over{2r_i^2 r_j^2}}=0}
\eqn\hmnpd{\dot\Phi=
{\ddot B_{ij}\over{\dot B_{ij}}}-\sum_i{\dot r_i\over{r_i}}}
\eqn\dild{2\sum_{i<j}{\dot r_i\dot r_j\over{r_i r_j}}+\dot\Phi^2-
\sum_{i<j} {\dot B_{ij}^2\over{2r_i^2 r_j^2}}=2\Lambda\ .}
To make contact with \metruu--\torst\ where the non-vanishing torsion
$B_{X_0 X_{D-2}}$ depends only on $t$,
we need consider only the values $i,j=0,1,2$.
The most general solution of the equations for the case of interest is thus
\eqn\ri{\eqalign{r_1^2
&={\Bigl(\sum_i{\alpha_i^2 \tan^{2p_i} \gamma t}\Bigr)\inv}\cr
  r_2^2&=r_1^2 A \tan^{2p_1+2p_2} \gamma t \cr
 B_{12}&=r_1^2\sum_i{B_i \tan^{2p_i} \gamma t }\cr
  \ee{\Phi}&=\ee{\Phi_0}{\dot B_{12}\over r_1 r_2}=
  \beta r_1^2\,\tan^{p_1+p_2-1} \gamma t\,\sec^2 \gamma t \ ,}}
where $ B_i$, $\Phi_0$, $\beta$, and $\alpha_i$ are arbitrary,  the $p_i$'s
are constrained as above, $A$ is given by $A\alpha_1 \alpha_2=\alpha_1^2
B_2-\alpha_2^2 B_1$, and the constant $\gamma$ remains as above.

To see that this is the most general
solution consistent with the ansatz is to count the number of independent
parameters. Equations \einstz--\hmnpd\ provide four second order
equations for the variables $r_1$, $r_2$, $\Phi$, $B_{12}$. This
allows eight free parameters given by the initial conditions
on the variables and their first derivatives. Equation \dild\
gives a nontrivial relation among them, which reduces the number to seven.
These seven parameters can be extracted from the expressions above,
taking into account a seventh parameter $t_0$ which results from the
freedom $t\to t+t_0$ to choose the origin of time (not an isometry).
Notice that
the symmetry $r_1\leftrightarrow r_2$ of the field equations \einstz--\dild\
is realized by the symmetries $p_i\to -p_i$ and $p_1\leftrightarrow p_2$ of the
circle described by the parameters $p_i$.
Again we can see that the expressions \metruu,\torsu\
obtained for the axial gauging are particular solutions of the above
equations for $p_1=1$, $p_2=0$, verifying that the WZW approach provides
solutions of the field equations to lowest order in $\alpha'$ if we shift
$k_0$ as before. Note also that we now have as well a solution for the
dilaton field.

This solution was only valid for one of the regions of the black hole
geometry ($0<uv<1$). It is easy to treat the other regions. For $uv<0$ we
can make the rotations $it\to R$, $r_2\to ir_1$ and $r_1\to r_2$. A
similar rotation, including the shift $t\to t+\pi/2$ gives the results for
region $uv>1$. There we see that the dilaton field, which gives the string
coupling constant, blows up as expected only at the singularity
$uv=1+\rho^2$. In the $D\to 26$ limit of these solutions, the torsion
vanishes and the solution collapses to the vector gauging case (making it
self-dual!). For $D>26$ analytic continuation gives the same picture as
for $D<26$ (as in the vector case), so we see that the critical dimension
plays an interesting role in our solutions. Note that we are unable to
obtain solutions for all allowed values of the $p_i$ parameters via a WZW
construction, but expect that exactly marginal deformations of the present
CFT's will access those parameters to complete the class of solutions.

\newsec{Duality}

The two different spacetime geometries, corresponding to the vector and
axial gaugings of the $G/H$ WZW model, can be viewed as different modular
invariant combinations of representations of the same holomorphic and
anti-holomorphic chiral algebras. There are general arguments \kir\ that
show that the vector and axial gaugings are dual, in the sense of having
equal partition functions. The duality is similar to the familiar $r\to
1/r$ duality in $c=1$ conformal field theory where two seemingly different
theories are as well related by a changing the sign of the left (or
holomorphic) currents $J=J_L$ with respect to the right (or
anti-holomorphic) currents $\overline J=J_R$. The lorentzian $D=2$ case is
special since the same geometry (2D black hole) is obtained by either the
vector or axial gauging, so we say that the model is self-dual. We now
point out the sense in which geometries for $D\geq 2$ are dual, placing
the vector/axial duality in a more generalized context.

In ref.~\busc, following previous developments
in supergravity, the $r\to 1/r$ duality of compactified string theories
was generalized to any string background for which the worldsheet action
has at least one isometry. For completeness, we review this analysis and
treat explicitly the case of $N$ commuting isometries in bosonic
string theory. The worldsheet action for the bosonic string is
\eqn\sigmau{S = {1\over{4\pi\alpha'}}\int \d^2z\, \Bigl(
\bigl(G_{MN}(X)+B_{MN}(X)\bigr)
\del X^M \delbar X^N + \alpha 'R^{(2)}\,\Phi (X)\Bigr)\ , }
where $M,N = 1,...D$; $G_{MN}, B_{MN}$ and $\Phi$ are the metric,
antisymmetric tensor and dilaton backgrounds respectively, and $R^{(2)}$ is
the 2D curvature. For a background with $N$ commuting isometries, we write the
action in the form
\eqn\sigmad{\eqalign{S = {1\over{4\pi\alpha'}}\int \d^2z\,
\Bigl(Q_{\mu\nu}(X_{\alpha})\,\del X^{\mu} \delbar X^{\nu}
 + Q_{\mu n}(X_{\alpha})\del X^{\mu} \delbar
X^n  + Q_{n \mu}(X_\alpha)\del X^n \delbar X^{\mu}
 \cr \qquad + Q_{mn}(X_\alpha)\del X^m \delbar X^n +
  \alpha' R^{(2)}\Phi(X_\alpha)\Bigr)\ ,\cr}}
where $Q_{MN}\equiv G_{MN}+B_{MN}$ and lower case latin indices $m,n$
label the isometry directions. Since the Lagrangian \sigmad\ depends on $X_m$
only through their derivatives, we can describe it in terms of the first order
variables $V^m=\del X^m$,
\eqn\sigmat{\eqalign{S = {1\over {4\pi\alpha'}}\int \d^2z\, \bigl(
  Q_{\mu \nu}(X_{\alpha})\,\del X^\mu \delbar X^{\nu}
+ Q_{\mu n}(X_{\alpha})\,\del X^{\mu} \overline V^n
 + Q_{n \mu}(X_{\alpha})\, V^n\delbar X^\mu
    \cr + Q_{mn}(X_{\alpha})\,V^m \overline V^n
+ \widehat X_m(\del \overline V^m - \delbar V^m)
+ \alpha' R^{(2)}\,\Phi (X_\alpha)\bigr)\ .\cr}}
This can be alternatively interpreted as gauging the isometry
with the constraint of vanishing gauge field strength \vero.
Integrating the Lagrange multipliers $\widehat X_m$ in the above
gives back \sigmad. After partial integration and solving for
$V^m$ and $\overline V^m$, we find the dual action
\eqn\dual{\eqalign{S' = {1\over {4\pi\alpha'}}\int &\d^2z\, \bigl(
Q'_{\mu \nu}(X_\alpha)\,\del X^\mu \delbar X^{\nu}
    +Q'_{\mu n}(X_\alpha)\,\del X^\mu\delbar\widehat X^n\cr
 &+Q'_{n \mu}(X_\alpha)\,\del \widehat X^n\delbar X^\mu
   +Q'_{mn}(X_\alpha)\,\del \widehat X^m\delbar \widehat X^n
    + \alpha'R^{(2)}\,\Phi'(X_\alpha) \bigr)\ .\cr}}
The dual backgrounds are given in terms of the original ones by
\eqn\qp{\eqalign{Q'_{mn}& = Q\inv_{mn}\cr
  Q'_{\mu \nu}& = Q_{\mu \nu} - Q\inv_{mn}\,Q_{n\nu}\,Q_{\mu m}\cr
  Q'_{n \mu}& = Q\inv_{nm}\, Q_{m \mu}\cr
  Q'_{\mu n}& = -Q\inv_{mn}\, Q_{\mu m}\ .\cr}}

To preserve conformal invariance, it can be seen by a careful
consideration of the path integral \busc\ (and from other approaches
\giva) that
\eqn\dil{\Phi' = \Phi - \log\sqrt{\det G_{mn} }}
Notice that equations \qp\ reduce to the usual duality transformations for
the toroidal compactifications of \grvsw\
in the limit $Q_{m\mu}=Q_{\mu m}=0$. For the case of a single isometry
($m=n=0$), we recover the
explicit expressions of \busc. This is to our knowledge the most general
statement of duality in string theory . In particular we see that a
space with no torsion ($Q_{m\mu} =Q_{\mu m}$) can be dual to a space with
torsion ($Q'_{m\mu}=-Q'_{\mu m}$), as found in the
previous section. To prove the duality we should
identify the particular isometry (of the $D+1$ total) that relates them.
Notice that for every isometry
we do not have to go to the first order formalism, i.e.\ we can
integrate the Lagrange multipliers $\widehat X_m$ for some of the
fields and instead the $V^m$ for the remaining fields with isometries.
This is the most general form of these duality transformations, and
eqs.~\qp\ and \dil\ should be read with indices $m,n$ running over only
the variables with isometries that have been dualized.

Before applying this formalism to the solutions we found in the previous
section, we compare this approach to duality with
others in the literature. In string theory, duality symmetry was
originally discovered in toroidal compactifications and found
to interchange winding states with momentum (Kaluza--Klein) states in the
compactified theory. We have seen that the toroidal compactification
is a particular case of a $\sigma$--model with isometries and thus
has this symmetry manifest. The interchange
of winding and momenta states realizes the duality symmetry in this
particular background but is not necessarily a generic feature of
duality, so we might expect duality even in backgrounds where
winding modes are not present. A particular example is given by the
2D Lorentzian black hole reviewed here in section 2.

The existence of duality as well implies
a continuous noncompact global symmetry \gz\ at the classical level
that relates the field equations of the theory \sigmad\ with the Bianchi
identities of the dual theory \dual.  For the case of two dimensional
$\sigma$--models in the context of string theory, this symmetry was found
in \cfg\ to be $SO(N,N)$.
Duality is of course a discrete subgroup of this continuous symmetry.
The noncompact continuous symmetries have been very useful to identify the
moduli space in certain string compactifications and more recently have
been used to find new nonstatic solutions from known ones
\refs{\gmv,\sen}. We wish to emphasize that they are not true string
symmetries since, just as in static backgrounds, they are broken by
nonperturbative effects on the worldsheet.
The surviving symmetry is a discrete symmetry which
includes duality and integer shifts of the antisymmetric tensor,
and as shown in \md\ generalizes to $SO(N,N,\IZ)$.

In order to make a connection between duality in this formulation and
the vector--axial duality
in $G/H$ WZW models, we recall the discussion of duality for the latter\kir.
If $H$ is abelian, group elements $g\in G$ can be parametrized as
$g=\ee{i\sigma\phi}\widehat g$ where $\sigma$
is the generator of $H$ (considered to be $U(1)$ for concreteness).
Substituting into the vector gauged WZW action, it turns out that the
action depends only on $\del\phi$, and proceeding with the standard
duality transformations \qp\
the axial gauged action is obtained. The isometry
in this case is then $\phi\to \phi + \delta\phi$ which is generated by the
right transformation $g\to gh$. After gauging
the $g\to hgh\inv$ transformation, we see that there will always be a remaining
isometry generated by the other independent transformation $g\to hg$, or more
symmetrically by the axial transformation $g\to hgh$. The same occurs for the
axial gauging, where the remaining isometry is given by the vector action
$g\to hgh\inv$.  (In the case of gauged non-abelian symmetries, it followed
from the analysis of section 2 that due to quantum effects the ungauged
symmetry does not remain even as a global symmetry.)

Now we are ready to analyze duality in the models of the previous section.
Starting with the vector gauging, we have to see how the gauged fixed
parameters transform under $g\to hg$, and go to a basis where only one of
the coordinates transforms. In that basis the metric is not necessarily
diagonal so the duality transformations \qp\ will be
non-trivial.\foot{Duality for the diagonal metric was explicitly proved in
\iltq\ to lowest order in $\alpha'$, and extended to next order in \bmq\
where either a field redefinition or equivalently a change in the dilaton
transformation was required. This had the
interesting consequence that large to small radius duality could be
explicitly realized in cosmology, as treated in \bv, but requiring only
weak coupling information from string theory, as recently
analyzed in \refs{\mv,\tv}. Duality for nondiagonal metrics relates not
only large to small radius, but as well relates different
cosmological models.} Let us consider the $0 < ab <1$ region for
concreteness. In that case we see that under $g\to hg$ the original
parameters transform as
\eqn\iso{\delta a=\varepsilon q_0  a\ ,\qquad
   \delta b=-\varepsilon q_0 b\ ,\qquad
   \delta r_i=\varepsilon q_i\ .}
We can see that the coordinates which diagonalize the metric transform as
\eqn\isod{\eqalign{&\delta t=0\cr
    &\delta X_0= \varepsilon (\rho^2+1) \cr
    &\delta X_i=0\ ,\qquad i= 1,\ldots, D-3\cr
    &\delta X_{D-2}=\varepsilon\ ,\cr}}
where we have used \ndnd.
Note that if we exchange $X_{D-2}$ for $Y\equiv X_0-\Omega X_{D-2}$,
with $\Omega\equiv\rho^2+1$,
the independent system of coordinates $t, X_0, Y$ and $X_i$
($i=1,\ldots, D-3$) is such that only $X_0$ transforms, so this defines
the isometry.

In these coordinates, the metric takes the form
\eqn\isot{\d s^2=-\d t^2+\Bigl({\tan^2 t+1\over\Omega^2}\Bigr) \d X^{2}_0
 +{1\over\Omega^2 }\,\d Y^2-{1\over \Omega^2 }\, \d X_0\, \d Y
 + \sum_{i=1}^{D-3} \d X_i^2}
{}From equations \qp, we can find the dual metric with the single isometry
$X_0\to X_0+\delta X_0$. It is straightforward to verify that it coincides
with the one given in equation \metruu\ coming from the $hgh$ gauging.
This proves that regions III of both geometries are mapped to each other
under duality. An identical analysis can be carried out for the other regions
obtained by analytic continuation of this region. It is straightforward
to see that region V of \ftdbh\ is mapped to region I of \ftdbs\
and in particular the singularity of the first is mapped to one horizon
in the second. Also, region I of the vector gauging black hole gets mapped
to regions V and VII together of the axial gauging black hole. This has
the interesting implication that a surface which in one geometry is perfectly
regular ($ab=\rho^2$) is mapped to the singularity in the
other geometry ($uv=1+\rho^2$). This goes even further than the black hole
singularity/horizon duality of the 2D black holes \dvv, since in that case
the horizon is a better behaved region than the singularity but
there remains nontrivial behavior such as the exchange of spacelike and
timelike coordinates. In the present case it can be seen explicitly that string
theory can deal with spacetimes that have singularities at the classical level,
in the sense that there still exists a description of interactions, etc.\ for
that region of spacetime. The situation is not so different from
situations have been encountered in compactified cases where singular spaces
(orbifolds) are dual to nonsingular ones (tori). (For a review in the simplest
$c=1$ case, see \rpglh; in a more general context that has recently arisen, see
e.g.\ \rvafa.) It would be interesting to study the
present geometries at the singularity in more detail to start probing
string theory in those regimes.

We have therefore established the duality between the
(2D black hole)$\otimes \IR^{D-2}$ and (3D black string)$\otimes \IR^{D-3}$
geometries. A more general analysis may be performed for the combination
of all the isometries using \qp\ and for the more general solutions
discussed earlier. Also we can easily check that the solutions \ri\ are
related to those of \mull\ for three dimensions by duality --- we rotate
the spacelike coordinates in \mull\ to get a nondiagonal metric and apply \qp.

Similarly,
starting from \mull\ for any number of dimensions we can find
new (cosmological) string solutions with torsion by applying \qp\ after going
to a nondiagonal basis. For the boosted variables,
\eqn\yi{ X_{M} \equiv \M_{MN} Y_N\ ,}
we see that dualizing equations \mull\ gives the metric (using the same
index convention mentioned after \qp )
\eqn\gtot{\eqalign{G_{mn}& =
\Bigl(\sum_{M}{\alpha _{M}^2 \tan ^{2p_M} \gamma t\,
\M_{Mm}\, \M_{Mn} }\Bigr)^{-1}\cr
G_{\mu \nu}& = G_{mn} \sum_{M,N}\alpha_ {M}^2\, \alpha_ {N}^2\,
    \tan ^{2p_M+2p_N} \gamma t \cr
   & \quad\cdot \M_{N\mu}\M_{Mn} (\M_{N\nu}\M_{Mm}-\M_{M\nu}\M_{Nm})\ ,\cr}}
and torsion
\eqn\torsn{B_{n\mu}=G_{nm} \sum_{M}\alpha_M
\tan^{2p_M}\gamma t \ \M_{Mm}\, \M_{M\mu}\ ,}
with a corresponding expression for the dilaton determined from \dil.
These expressions provide new explicit cosmological solutions with torsion
for any value of $m,n\le N$ (and $N$ varying from one to the total number
of isometries). Notice that they depend on $D^2-D+1$ arbitrary parameters
($\Gamma_{MN}\equiv \alpha_M \M_{MN}$, $p_M$, and $t_0$) which equals the
number of boundary conditions allowed for $G_{MN} , B_{MN}$ and $\Phi $
and their first derivatives (minus the constraint provided by equation
\dil). Eq.~\ri\ is the $D=3$ case of these general solutions. It is
certainly interesting to explore the possible cosmological consequences of
these solutions as well as the implications of their duality to the known
solutions of \muller. Similar remarks apply for the generalization of the
other black hole regions (by analytic continuation) and it would be
interesting to determine if together they could lead to a geodesically complete
system of coordinates for a black hole type of geometry with \gtot\
as the cosmological region. This is certainly true for the cases we
constructed in section 3 ($p_1=1$, all other $p_i=0$), where
\gtot\ and \torsn\ generate new dual black branes (although they are not
directly obtained from the WZW construction).

\newsec{Conclusions}

We have presented a general discussion of the class of geometries
that can be obtained in string theory from noncompact coset conformal
field theories in the large $k$ limit. The WZW approach plays a crucial
role in allowing identification the background fields, in
particular the metric of the target spacetime. This is not
necessary for the CFT describing the internal degrees of freedom
in superstring compactifications, since in that case a geometrical
interpretation for that sector of the string vacua is unnecessary
(and does not even necessarily exist in general).

We have found that the number of possible geometries obtained in this way
is very restricted due to the constraint of having a single timelike
coordinate.\foot{Euclidean cosets, corresponding to the cases listed in
table 1, remain useful for describing instanton-like configurations such
as euclidean black holes, and for describing compactified dimensions.} In
the case of superstring compactifications, however, the coset $G/H$
describing the internal degrees of freedom actually provides many string
vacua. These arise both from the different possible choices of boundary
conditions (orbifoldizing) and also because they are only special points
in a degenerate space of vacua parametrized by the exactly marginal
deformations of the CFT. In the case of (2,2) compactifications, for
example, the coset models describe particular points of a Calabi-Yau
manifold.

The study of the spectrum of the noncompact coset models, especially the
marginal deformations, will thus generate new classes of geometries which
would be interesting to investigate and still represent exact CFT's. A
motivation for considering the models of section 3 was to find a conformal
field theory realization of the cosmological solutions found in \muller.
We have only partially succeeded since in the vector gauging our solutions
in the cosmological region correspond to those of \muller\ only for fixed
values of the parameters $p_i$ of equations \mull. A natural expectation
is that exactly marginal deformations of our models will turn on those
parameters $p_i$ and generate the whole class of cosmological models of
reference \muller, as well as the new class of models with torsion we give
in \ri\ for the axial gauging. This could be an interesting extension of
the results here.
Another possibility is that by marginal deformations, the black
hole--like singularities of the noncompact cosets could be ``blown-up'',
reminiscent of the way orbifold singularities can be blown-up to obtain
smooth string compactifications. This would illustrate the ameliorating
control that string theory seems to have over singularities, already
exemplified in section 4 by the duality between singular and regular
regions of the different black hole geometries of section 3.

Our discussion of duality in section 4, following \busc, is based
entirely on the existence of isometries of the $\sigma$--model.
Although it is the most general statement of that symmetry to date
(including the known toroidal compactifications as particular cases),
it cannot be the final statement because we know that there are geometries,
such as Calabi-Yau spaces, that have no continuous isometries and still
are known to have duality-like symmetries, for example the mirror symmetry
of \px. It would be very interesting to have a
unified understanding of these dualities from a $\sigma$--model
point of view.

Finally, since the subject of the present article has been evolving faster
than our ability to write it up, we briefly discuss some of
those recent results that have partial overlap with our work. First, the
list of single-time cosets for the case of simple groups (table 2) was
independently given in \barst\ using the known list of supersymmetric
(K\"ahler) cosets (with the exception of the $SO(D-1,2)/SO(D-1,1)$ given
earlier in \bn) but with no claim to completeness.
The particular case $D=3$ of the axial gauging geometry of the models of
section 3, was discussed in \hh\ where a complete discussion of
the black string geometry can be found. The expressions for the background
fields given therein are in agreement with those given here after
a simple change of variables. Duality of that geometry
 was discussed in \hhs\ and the relation with the vector gauging was
briefly mentioned, although the proof of the duality of both geometries
was not explicitly presented.\foot{In \hh, duality is found to
map the singularity of the axial gauged geometry to a
region inside the singularity of the vector gauged geometry, instead of to the
asymptotically flat region as found here. The discrepancy is due to the
presence of two isometries, and hence more than one possible duality.
The two results are easily reconciled by applying a second duality
transformation.} More recently, duality for several commuting
symmetries was discussed in \giro\ and an $SO(N,N,\IZ)$ was identified as
the modular group in agreement with our comments about the general results
of \cfg\ and \md.
The periodic coordinates and the winding mode
/ momentum duality insisted upon by these authors, however, is not considered
essential here. The solution of 3D
cosmological backgrounds discussed in \ind\ are particular cases of our
solutions \ri\ (up to analytic continuations). Other models have
been recently explored \cresc\ and together with the models
considered here, most of the possible cases of our table 3 for $D\le 4$ have
been investigated.


\bigbreak\bigskip\bigskip\centerline{{\bf Acknowledgements}}\nobreak
We thank I. Bars, J. Distler, and E. Verlinde for discussions at the Aspen
Center for Physics. We are also grateful for helpful conversations with P.
Candelas, X. de la Ossa, J.-P. Derendinger, L. Ib\'a\~nez, M.
Crescimanno, and as well as J. Horne and G. Horowitz for information on their
related work. This work was supported in part by DOE contract W-7405-ENG-36.

\listrefs


\vfuzz=2pt\baselineskip16pt\nopagenumbers
\def\Captionbox#1#2{\vbox{\hbox{#1}\break\vphantom{x}\break\hbox{#2}}}

$$\vbox{\offinterlineskip\halign{\strut#&
       \vrule height10.4pt depth5.4pt#&\hfil$\,#\,$\hfil&
       \vrule#&\hfil$\,#\,$\hfil&
       \vrule#&\hfil$\,#\,$\hfil&
       \vrule#&\hfil$\,#\,$\hfil&
       \vrule#\cr
      \noalign{\hrule}
&&G&&H&&\# $H$  {\rm\ Generators} &&{\rm Dim}$(G/H)$&\cr
\noalign{\hrule\vskip3pt\hrule}

&&SL(p,\IC)&& SU(p)&& p^2-1&& p^2-1&\cr\noalign{\hrule}
&&SL(p,\IR)&& SO(p)&& {1\over 2}p(p-1)&&
  {1\over 2}p(p+1)-1&\cr\noalign{\hrule}
&&SU^*(2p)&& USp(2p) && p(2p+1) &&  (p-1)(2p+1)&\cr\noalign{\hrule}
&&SU(p,q)&& SU(p)\times SU(q)\times U(1)&& p^2+q^2-1
 && 2pq&\cr\noalign{\hrule}
&&SO(p,\IC) && SO(p,\IR) && {1\over 2}p(p-1) &&
 {1\over 2} p(p-1)&\cr\noalign{\hrule}
&&SO(p,q) &&SO(p)\times SO(q)&& {1\over 2}\bigl(p(p-1)+q(q-1)\bigr) &&
 pq&\cr\noalign{\hrule}
&&SO^*(2p) && SU(p)\times U(1) && p^2  && p(p-1)
&\cr\noalign{\hrule}
&&Sp(2p,\IC) && USp(2p)  && p(2p+1) &&
p(2p+1)&\cr\noalign{\hrule}
&&Sp(2p,\IR)&& SU(p)\times U(1) && p^2 && p(p+1) &\cr\noalign{\hrule}
&&USp(2p,2q) &&USp(2p)\times USp(2q)&& p(2p+1)+q(2q+1)&&
4pq&\cr\noalign{\hrule}
&&G_2^c&&G_{2 (-14)}&&14&& 14&\cr\noalign{\hrule}
&&G_{2 (+2)} && SU(2)\times SU(2) &&6  &&  8&\cr\noalign{\hrule}
&&F_4^c && F_{4(-52)} && 52 && 52&\cr\noalign{\hrule}
&&F_{4 (+4)} && USp(6)\times SU(2) && 24 && 28&\cr\noalign{\hrule}
&&F_{4 (-26)} && SO(9)&& 36 &&  16&\cr\noalign{\hrule}
&&E_{6}^c && E_{6 (-78)} && 78 &&  78&\cr\noalign{\hrule}
&&E_{6 (+6)} && USp(8) && 36 && 42&\cr\noalign{\hrule}
&&E_{6(+2)} && SU(6)\times SU(2) && 38 &&  40&\cr\noalign{\hrule}
&&E_{6 (-14)} && SO(10)\times SO(2) && 46 && 32&\cr\noalign{\hrule}
&&E_{6 (-26)}&& F_{4 (-52)}&& 52 &&  26&\cr\noalign{\hrule}
&&E_{7}^c && E_{7 (-133)} && 133 &&  133&\cr\noalign{\hrule}
&&E_{7(+7)} && SU(8) && 63 && 70&\cr\noalign{\hrule}
&&E_{7 (-5)} && SO(12)\times SO(3) && 69 &&  64&\cr\noalign{\hrule}
&&E_{7 (-25)} && E_{6 (-78)}\times SO(2) && 79 &&  54&\cr\noalign{\hrule}
&&E_{8}^c && E_{8 (-248)} && 248 &&  248&\cr\noalign{\hrule}
&&E_{8 (+8)} && SO(16) && 120 &&  128&\cr\noalign{\hrule}
&&E_{8 (-24)} && E_{7 (-133)} && 136   && 112&\cr\noalign{\hrule}
}}$$

\medskip\nobreak
\vbox{\noindent Table 1:
Noncompact coset spaces $G/H$ with no timelike coordinates. $G$ is simple and
$H$ is the maximal compact subgroup.}
\vfill\eject
\vfill\eject\ifx\answ\bigans\else\quad\vfill\eject\fi
$$\vbox{\offinterlineskip\halign{\strut#&
       \vrule height15pt depth10pt#&\hfil$\,#\,$\hfil&
       \vrule#&\hfil$\,#\,$\hfil&
       \vrule#&\hfil$\,#\,$\hfil&
       \vrule#&\hfil$\,#\,$\hfil&
       \vrule#&\hfil$\,#\,$\hfil&
       \vrule#&\hfil$\,#\,$\hfil&
       \vrule#&\hfil$\,#\,$\hfil& \vrule#\cr
      \noalign{\hrule}
&&G&&H&&\multispan 3 \hfil\#
 $G$ Generators\hfil &&\multispan 3 \hfil\#
 $H$ Generators\hfil&&{\rm Signature}&\cr
      \noalign{\hrule}
&&{}&&{}&&{\rm compact}&&
$\raise -5pt\Captionbox{\kern 10pt non}{compact}$
&&{\rm compact}&&$\raise -5pt\Captionbox{\kern 10pt  non}{compact}$&&{}&\cr
      \noalign{\hrule\vskip3pt\hrule}
&&SU(p,q)&&SU(p)\times SU(q)&&p^2+q^2-1&&2pq&&p^2+q^2-2&&0&&(1,2pq)&\cr
      \noalign{\hrule}

&&SO(p,2)&&SO(p,1)&&{1\over 2}p(p-1)+1&&2p&&{1\over 2}p(p-1)&&
p&&(1,p)&\cr
\noalign{\hrule}
&&SO(p,2)&&SO(p)&&{1\over 2}p(p-1)+1&&2p&&{1\over 2}p(p-1)&&0&&
(1,2p)&\cr\noalign{\hrule}
&&Sp(2p,\IR)&&SU(p)&&p^2&&p(p+1)&&p^2-1&&0&&(1,p(p+1))&\cr\noalign{\hrule}
&&SO^*(2p)&&SU(p)&&p^2&&p(p-1)&&p^2-1&&0&&(1,p(p-1))&\cr\noalign{\hrule}
&&E_{6 (-14)}&&SO(10)&&46&&32&&45&&0&&(1,32)&\cr\noalign{\hrule}
&&E_{7 (-25)}&&E_{6 (-78)}&& 79 && 54 &&78 && 0
&&(1,54)&\cr\noalign{\hrule}}}$$
\bigskip
\centerline{Table 2: Coset spaces $G/H$
with only one time coordinate (for simple groups $G$)}

\def\inbar{\vrule height1ex width.4pt depth0pt}
\def\IC{\relax\thinspace\inbar\kern-.3em{\rm C}}
\vfill\eject\ifx\answ\bigans\else\quad\vfill\eject\fi

\null\vskip-60pt
\def\ucaptionbox#1{$$\vbox{{}\break\vphantom{x}\hbox{#1}\break\vir}$$}
\def\dcaptionbox#1#2{$$\vbox{{}\break\vphantom{x}\hbox{#1}\break\vphantom{x}
\break\hbox{#2}\break\vir}$$}
\def\tcaptionbox#1#2#3{$$\vbox{{}\break\vphantom{x}\hbox{#1}\break\vphantom{x}
\break\hbox{#2}\break
 \vphantom{x}\break\hbox{#3}\break\vir}$$}
\def\ccaptionbox#1#2#3#4{$$\vbox{{}\break\vphantom{x}%
\hbox{#1}\break\vphantom{x}
\break\hbox{#2}\break
 \vphantom{x}\break\hbox{#3}\break\vphantom{x}\break\hbox{#4}\break\vir}$$}
\def\virt#1{\vrule height #1 width 0pt}
\def\vir{\hbox{\virt{3pt}}}
$$\vbox{\offinterlineskip\halign{\strut#&
       \vrule#&\hfil$\,#\,$\hfil&
       \vrule#&~~$#$~~\hfil&
       \vrule#\cr
       \noalign{\hrule}
&&\hfil$\ucaptionbox{D}$
&&\hfil$\ucaptionbox{G/H}$& \cr
\noalign{\hrule\vskip3pt\hrule}
&&\dcaptionbox{2}{\virt{1.5pt}}
&&$\ucaptionbox{${SL(2,\IR)\over SO(1,1)}$}$&\cr
\noalign{\hrule}
&&\dcaptionbox{3}{\virt{0.1pt}}
&&$\ucaptionbox{${SO(2,2)\over SO(2,1)};\ \{ (D=2{\rm\ case})  ;\
                           {SL(2,\IR)\over U(1)}
                           \} \times \IR$}$&\cr\noalign{\hrule}

&&\dcaptionbox{4}{\virt{0.1pt}}
&&$\ucaptionbox{${SO(3,2)\over SO(3,1)} ;\  {SO(2,2)\over SO(1,1)
                            \times SO(2)};\  
	  {SO(3,\IC)\over SO(3,\IR)} ;\ {SO(3,1)\over SO(3)} ;\
             (D=3) \} \times\IR$}$&\cr\noalign{\hrule}

&&\dcaptionbox{5}{\virt{10pt}}
&&$\dcaptionbox{${SU(2,1)\over SU(2)}$;\  ${SO(4,2)\over SO(4,1)}$;\
     ${SO(2,2)\over SO(2)}$;\
     ${SL(2,\IR)\times SL(2,\IC)\over SO(1,1)\times SU(2)}$}
                       {$\{ {SU(2,1)\over SU(2)\times U(1)};\
   {SO(2,2)\over SO(2)^2};\     {SO(4,1)\over SO(4)};\
                   (D=4)\}\times \IR$}$&\cr\noalign{\hrule}
&&\dcaptionbox{6}{\virt{10pt}}
&& $\dcaptionbox{${SO(5,2)\over SO(5,1)}$;\
 $\{ {SL(3,\IR)\over SO(3)} $;\
       ${SO(5,1)\over SO(5)}$;\ $(D=5)\} \times \IR$;\
        $(D=4) \times {SL(2,\IR)\over SO(2)}$}
  {$(D=3) \times {SL(2,\IC)\over SU(2)}$;\ $(D=2) \times \{
  {SU(2,1)\over SU(2)\times U(1)}$;\
   ${SO(4,1)\over SO(4)}$;\ ${SO(2,2)\over SO(2)^2 } \}$ }$&\cr\noalign{\hrule}
&&\dcaptionbox{7}{\virt{20pt}}
&& $\tcaptionbox{$\{ {SU(3,1)\over SU(3)\times U(1)};\
       {SO(4,\IC)\over SO(4,\IR)};\ {SO(6,1)\over SO(6)};\
       {SO(3,2)\over SO(3)\times SO(2)};\ (D=6) \}\times \IR$}
        {$ {SU(3,1)\over SU(3)};\  {SO(6,2)\over SO(6,1)};\ {SO(3,2)\over
SO(3)}
;\  (D=5) \times {SL(2,\IR)\over SO(2)};\ (D=4) \times {SL(2,\IC)\over
        SU(2)}$} {$(D=3)\times \{ {SU(2,1)\over SU(2)\times U(1)};\
        {SO(4,1)\over SO(4)};\ {SO(2,2)\over SO(2)^2}\};\
        (D=2)\times \{ {SL(3,\IR)\over SO(3)};\
       {SO(5,1)\over SO(5)} \}$}$&\cr\noalign{\hrule}
&&\dcaptionbox{8}{\virt{22pt}}
&& $\tcaptionbox{${SO(7,2)\over SO(7,1)};\ \{  {SO(7,1)\over SO(7)};\
    (D=7) \}\times \IR;\ (D=6)\times{SL(2,\IR)\over SO(2)};\
	  (D=5)\times {SL(2,\IC)\over SU(2)}$}
         {$(D=4)\times \{ {SU(2,1)\over SU(2)\times U(1)};\
        {SO(4,1)\over SO(4)};\ {SO(2,2)\over SO(2)^2}\};\
	(D=3)\times \{ {SL(3,\IR)\over SO(3)};\
       {SO(5,1)\over SO(5)} \}$}
       {$(D=2)\times\{{SU(3,1)\over SU(3)\times U(1)};\
       {SO(4,\IC)\over SO(4,\IR)};\ {SO(6,1)\over SO(6)};\
       {SO(3,2)\over SO(3)\times SO(2)}\}$}$&\cr\noalign{\hrule}
&&\dcaptionbox{9}{\virt{30pt}}
&& $\ccaptionbox{${SU(2,2)\over SU(2)\times SU(2)};\
                {SU(4,1)\over SU(4)};\  {SO(8,2)\over SO(8,1)};\
                {SO(4,2)\over SO(4)};\ (D=7)\times {SL(2,\IR)\over SO(2)};\
                (D=6)\times{SL(2,\IC)\over SU(2)}$}
                {$(D=5)\times \{ {SU(2,1)\over SU(2)\times U(1)};\
        {SO(4,1)\over SO(4)};\ {SO(2,2)\over SO(2)^2}\};\ (D=4)\times \{
        {SL(3,\IR)\over SO(3)};\ {SO(5,1)\over SO(5)} \}$}
        {$(D=3)\times\{{SU(3,1)\over SU(3)\times U(1)};\
       {SO(4,\IC)\over SO(4,\IR)};\ {SO(6,1)\over SO(6)};\
       {SO(3,2)\over SO(3)\times SO(2)}\};\ (D=2)\times {SO(7,1)\over SO(7)};\
       \IR\times$}
         {$\{{SL(3,\IC)\over SU(3)};\
                {SU(2,2)\over SU(2)^2\times U(1)};\
                {SU(4,1)\over SU(4)\times U(1)};\
                 {SO(8,1)\over SO(8)};\
                {SO(4,2)\over SO(4)\times SO(2)};\
                {USp(4,2)\over USp(4)\times USp(2)};\
              (D=8)\}$}$&\cr\noalign{\hrule}
&&\dcaptionbox{10}{\virt{30pt}}
&& $\ccaptionbox{$ (D=8)\times {SL(2,\IR)\over SO(2)};\
                (D=7)\times{SL(2,\IC)\over SU(2)};\
                (D=6)\times \{ {SU(2,1)\over SU(2)\times U(1)};\
        {SO(4,1)\over SO(4)};\ {SO(2,2)\over SO(2)^2}\}$}
        {$(D=3)\times {SO(7,1)\over SO(7)};\
        (D=4)\times\{{SU(3,1)\over SU(3)\times U(1)};\
       {SO(4,\IC)\over SO(4,\IR)};\ {SO(6,1)\over SO(6)};\
       {SO(3,2)\over SO(3)\times SO(2)}\}$}
       {$ (D=2)\times \{
       {SL(3,\IC)\over SU(3)};\ {SU(2,2)\over SU(2)\times U(1)};\
       {SU(4,1)\over SU(4)\times U(1)};\
                 {SO(8,1)\over SO(8)};\
                {SO(4,2)\over SO(4)\times SO(2)};\
                {USp(4,2)\over USp(4)\times USp(2)}\}$}
        {$(D=5)\times \{
        {SL(3,\IR)\over SO(3)};\ {SO(5,1)\over SO(5)} \};\
	\{{SL(4,\IR)\over SO(4)};\
                 {SO(9,1)\over SO(9)};\
       (D=9)\}\times\IR;\ {SO(9,2)\over SO(9,1)}$}$&\cr\noalign{\hrule}
}}$$
\nobreak
\noindent \vbox{\baselineskip 14pt \noindent Table 3:
Noncompact coset spaces $G/H$ with one time coordinate with
dim$(G/H)\leq 10$, where $G$ is a product of simple noncompact groups.}

\bye